\documentclass[a4paper,11pt]{article}
\pdfoutput=1 

\usepackage{jheppub} 

\usepackage[T1]{fontenc} 
\usepackage{amsmath,amssymb}
\usepackage{dsfont}
\usepackage{upgreek}
\usepackage{slashed}
\usepackage{appendix}
\usepackage{tabu}
\newcommand{\be}{\begin{equation}}
\newcommand{\ee}{\end{equation}}
\newcommand{\ben}{\begin{eqnarray}}
\newcommand{\een}{\end{eqnarray}}

\def\vo{\mathcal{V}}

\title{\bf{Moduli Stabilisation with Nilpotent Goldstino: Vacuum Structure and SUSY Breaking}}
 \author[a]{\small{Luis Aparicio},}
\author[a,b]{\small{Fernando Quevedo},}
\author[c,d,a]{\small{Roberto Valandro}}
\affiliation[a]{\small{ICTP, Strada Costiera 11, 34151 Trieste, Italy}}
\affiliation[b]{\small{DAMTP, CMS, University of Cambridge, Wilberforce Road, Cambridge, CB3 0WA, UK.}}
\affiliation[c]{\small{Dipartimento di Fisica dell'Universit\`a di Trieste,
Strada Costiera 11, 34151 Trieste, Italy.
}}
\affiliation[d]{\small{INFN, Sezione di Trieste, , Via Valerio 2, 34127 Trieste, Italy.
}}
\emailAdd{laparici@ictp.it}
\emailAdd{f.quevedo@damtp.cam.ac.uk}
\emailAdd{roberto.valandro@ts.infn.it}

\abstract{\small{We study the effective field theory of KKLT and LVS moduli
stabilisation scenarios coupled to an anti-D3-brane at the tip of a warped
throat.
We describe the presence of the
anti-brane in terms of a nilpotent goldstino superfield in a supersymmetric
effective field theory. The introduction of this superfield
produces a term that can lead to a de Sitter minimum.
We fix the K\"ahler moduli dependence of the nilpotent field couplings by
matching this term with the anti-D3-brane uplifting contribution. The main result of this paper is the computation, within this EFT, of the soft supersymmetry breaking terms in both KKLT and
LVS for matter living on D3-brane (leaving the D7-brane analysis to an
appendix). A handful of distinct
phenomenological scenarios emerge that could have low energy
implications, most of them having a split spectrum of soft masses. Some cosmological and phenomenological properties of these models are discussed.
We also check that the attraction
between the D3-brane and the anti-D3-brane does not affect the leading
contribution to the soft masses and does not destabilise the system.
}}

\preprint{DAMTP-2015-64} 

\begin{document} 
\maketitle
\flushbottom

\section{Introduction}\label{Sec:Intro}

Constrained superfields can play an important role in supersymmetric theories and have been subject to intensive research during the past few years. The simplest case is the  nilpotent chiral superfield $X$  ($X^2=0$) (see for instance \cite{rocek,Ivanov:1978mx,lindstrom,casalbuoni, Komargodski:2009rz} and references therein). $X$ has a single propagating component, the Volkov-Akulov goldstino \cite{Volkov:1973ix}, and  supersymmetry (susy) is broken by its F-term. 
The supersymmetry is realised nonlinearly, but it can nevertheless be represented by the standard supersymmetric couplings of  chiral, gauge and gravity  superfields coupled to the goldstino superfield $X$. Implementing this idea into the low energy effective action of string compactifications allows to describe the presence of the anti-brane by using a supersymmetric action. 

In type IIB flux compactifications \cite{gkp,Dasgupta:1999ss,Grana:2005jc,Douglas:2006es,Blumenhagen:2006ci}\, the presence of an anti-D3-brane, as proposed by Kachru, Kallosh, Linde and Trivedi (KKLT) in \cite{kklt}, provides probably the simplest and more model independent realisation of de Sitter (dS) space in string theory.\footnote{See \cite{Burgess:2003ic,Balasubramanian:2004uy,Villadoro:2005yq, Lebedev:2006qq, Dudas:2006gr, Westphal:2006tn,deAlwis:2011dp,Rummel:2011cd,bcmq,Rummel:2014raa} for other mechanisms to find dS vacua in the EFT of type IIB flux vacua and \cite{Cicoli:2012vw,Louis:2012nb,Cicoli:2013mpa,Cicoli:2013cha,Braun:2015pza} for explicit realisations in concrete models.} 
The anti-brane breaks the supersymmetry preserved by the rest of the compactification background at the (warped) string scale. In the original paper, its positive contribution to the supergravity scalar potential was simply added as an explicitly supersymmetry breaking term to the supergravity effective action.
For this reason, the control over such non-supersymmetric effective field theory was questioned.\footnote{In the last years a debate on the (meta)stability  of this setup was raised by \cite{DeWolfe:2008zy,McGuirk:2009xx,Bena:2009xk}. Recent development can be found in
\cite{Bena:2014jaa,Danielsson:2014yga,Michel:2014lva,Hartnett:2015oda,Bena:2015kia,Cohen-Maldonado:2015ssa,Bertolini:2015hua,Polchinski:2015bea,Cohen-Maldonado:2015lyb}.}
Describing the effective field theory (EFT) that captures the physics of this anti-brane in terms of a purely supersymmetric formulation is then highly desirable. 
Recently there has been progress in this direction. The effective  couplings were studied in \cite{Farakos:2013ih,adfs,sugra, sugramore,Antoniadis:2015ala,Hasegawa:2015bza,Kallosh:2015tea,Zwirner,Shillo,Kallosh:2015pho} and the KKLT uplifting term was reproduced in the supergravity framework in \cite{kw,kw2,kqu}. Finally, in \cite{kqu} explicit string constructions were presented in which an anti-D3-brane at the tip of a flux-induced throat has only the goldstino as its massless degree of freedom,  justifying the use of the nilpotent field $X$ to describe the anti-brane.\footnote{
Nonlinear supersymmetry in string theory was also discussed in \cite{Dudas:2000nv,Pradisi:2001yv}.
}
A complementary approach has been recently presented in \cite{Bandos:2015xnf}, 
where the authors introduce a
locally supersymmetric generalisation of the Volkov-Akulov goldstino action that describes a non-BPS D3-brane in  superspace
and couple it to the minimal $N=1$ 4D supergravity.

Over the past decade much work has been dedicated to the effective field theory of moduli stabilised de Sitter vacua. Both the KKLT \cite{kklt} and Large Volume (LVS) \cite{Balasubramanian:2005zx} scenarios have been explored in order to extract the low energy properties of chiral matter fields.  It is the purpose of this article to revisit and compute the soft breaking terms induced by the presence of the nilpotent superfield $X$. We recall that even though in KKLT the anti-D3-brane is the source of supersymmetry breaking, in LVS the anti de Sitter (AdS) minimum is already non-supersymmetric with the F-term of the volume modulus providing the main source of supersymmetry breaking. It is anyway desirable to have all the sources of supersymmetry breaking described in terms of the same supergravity effective field theory. 

Before studying the soft susy breaking terms, 
we briefly review the properties of the nilpotent goldstino superfield $X$ and its couplings in the supergravity effective action in Section \ref{Sec:NilpGold}. We constrain the moduli dependence of its contribution to the K\"ahler potential and to the superpotential, by matching the generated uplift term to the scalar potential with the one generated by an anti-D3-brane. We found that a compact logarithmic no-scale form of the K\"ahler potential is in principle possible if one relation between the coefficients of the relevant terms holds. 
Section \ref{Sec:WarpFlComp} is devoted to describe the physics of string compactifications with both D3 and  anti-D3-branes. We briefly recall the discussion of warped compactifications with  mobile D3-branes presented in \cite{kklmmt} by Kachru, Kallosh, Linde, Maldacena, McAllister and Trivedi (KKLMMT). We are also able to reproduce the brane/anti-brane Coulomb interactions by
adding a coupling between $X$ and the D3-brane moduli in the superpotential.
We finally estimate the potential instability due to the brane/anti-brane attraction and find that this is usually too weak to compete with the generic magnitude of soft scalar masses (that stabilise the D3-brane position). However for the case of ultralocal scalar masses discussed in \cite{bckmq, ackmmq} there are several cancellations and the soft masses become of order ${\mathcal{O}}(m_{3/2}/\vo)$. We find that this value is at the border of the stability bound. 

In Sections \ref{Sec:KKLT} and \ref{Sec:LVS} we study susy breaking for KKLT and LVS. 
In Section \ref{Sec:KKLT} we extend the analysis started in \cite{kqu}, where the soft breaking terms in KKLT were computed by using the nilpotent superfield formalism.  Generically the spectrum is split in the sense that scalar masses are a few orders of magnitude larger than gaugino masses. This is not the case for D3-branes at singularities if the K\"ahler potential can be put in the compact logarithmic form we mentioned above. 
When this happens, there are cancellations in the scalar masses that make the subleading contributions relevant. The most important ones come from leading order $\alpha'$ corrections. Including them, we find the non-vanishing values of scalar masses and we compare them with those coming from anomaly mediation which at this level can be competitive. 
The resulting structure of soft terms for the KKLT case has some analogies with the one originally found using other techniques by \cite{cfno,cfno2,cfno3,Lebedev:2006qq}.  In particular, the form of the leading contribution of the scalar masses is very similar: it is proportional to the gravitino mass and the overall factor depends on the K\"ahler metric of the matter fields and of the chiral multiplet responsible for the uplift.  Moreover, for a particular value of this coefficient, the leading contribution is zero and the subleading terms become important.
There are however some differences: first of all, in our case the (uplifting) chiral superfield is constrained (nilpotent) and it is claimed to have a specific origin (it describes the only degree of freedom of an anti-D3-brane on top of an orientifold O3-plane in presence of three-form fluxes); moreover, its K\"ahler metric is different with respect to the one used in the original paper. 
Second, we relate the possible vanishing of the leading term in the scalar masses to a conjectured logarithmic form of the K\"ahler potential (in analogy with how the D3-brane fields behave).
Third, we consider more effects contributing to the subleading terms in the soft masses, i.e. both non-perturbative and $\alpha'$ corrections and anomaly mediation contributions, that can compete to avoid or force tachyonic masses.

In Section \ref{Sec:LVS} we revisit LVS soft breaking terms that have been recently studied in \cite{ackmmq} using different sources of uplifting. We consider the sequestered scenario (like in \cite{bckmq,ackmmq})  in which the uplifting mechanism is relevant. We found a very similar soft term structure. In Section~\ref{Sec:CosmoPhenoObserv} we outline the phenomenological and cosmological implications of each of the scenarios described in sections \ref{Sec:KKLT} and \ref{Sec:LVS}. In Section \ref{Concl} we conclude. In the appendices we  discuss the soft terms for D7-branes and the anomaly mediation effects.

\section{Effective field theory of the nilpotent goldstino}\label{Sec:NilpGold}

When supersymmetry is spontaneously broken in supergravity effective theories,
the goldstino is eaten by the gravitino realising the super-Higgs effect. 
If this breaking happens at low energies compared with the Planck mass, the goldstino couplings can be described by introducing 
a (constrained) independent superfield in the supergravity effective action. This has a non-linearly realised supersymmetry,
as in the original Volkov-Akulov formalism. 

Sometimes the process of supersymmetry breaking is not fully under control, like for example in situations with
strongly coupled systems or in D-brane models in which the presence of different objects can break supersymmetry (sometimes even partially). It is nevertheless important to have control on the low energy effective theory in which supersymmetry is non-linearly realised. Over the years there have been several approaches to describe the low energy couplings of the goldstino in terms of spurion or constrained superfields (see for instance \cite{Komargodski:2009rz} and references therein).  We will describe the goldstino in terms of a chiral superfield $X$ that is further constrained to be nilpotent, i.e. $X^2=0$. This has been claimed to be the right approach to deal with the breaking of supersymmetry induced by the presence of an anti-D3-brane in flux compactifications \cite{kw,kw2,kqu}.

The couplings of a nilpotent chiral superfield can be described in terms of very simple K\"ahler potential  $K$, superpotential $W$ and gauge kinetic function $f$ as follows:
\be\label{EFTXVA}
K=K_0  + K_1X + \bar{K}_1\bar{X}\ + K_2X\bar{X} , \qquad W=\rho X + W_0 , \qquad f=f_0+f_1 X,
\ee
where $K_0, K_1, K_2, \rho, W_0, f_0, f_1$ may be functions of other low energy fields. Higher powers of $X$ are not present in $K$ and $W$ because $X^2=0$.

Furthermore the nilpotency condition implies a constraint on the components of the chiral superfield $X$, where
\be
X=X_0(y)+\sqrt{2}\psi(y) \theta + F(y)\theta\bar{\theta} \:,
\ee
with, as usual, $y^\mu=x^\mu+ i \theta \sigma^\mu \bar{\theta}$. In fact, imposing $X^2=0$  implies 
\be
X_0=\frac{\psi\psi}{2F}\:.
\ee

The effective field theory of $X$ with K\"ahler and superpotential  \eqref{EFTXVA} reproduces the Volkov-Akulov action, that has been studied both in global and local supersymmetry. For the anti-D3-brane in the KKLT scenario, the representation in terms of $X$ is very convenient since it allows to treat its effect in terms of standard supergravity couplings of matter and moduli superfields to the nilpotent goldstino. 

Recently, it has been shown that the nilpotent superfield is enough to capture all the anti-D3-brane degrees of freedom when this brane is placed on top of an orientifold plane 
\cite{kqu}: a combination of fluxes and orientifold projections leave  the massless goldstino as the only low energy propagating particle, thus justifying the use of a nilpotent superfield $X$ to account for the presence of the anti-brane in the low energy effective field theory. 
The simplest example is when an O3-plane and an anti-D3-brane are at the tip of the warped throat.
In this case, the anti-D3-brane 
does not have a modulus describing its motion, contrary to D3-branes in the bulk. This fits with the fact that the scalar component of $X$ is not a propagating field.
Moreover, in calculating the scalar potential, there is no contribution from $X_0$ and it is consistently set to zero when looking for Lorentz preserving vacuum configurations as we set all fermions to zero. This simplifies substantially the calculations.

Let us consider the couplings of $X$ with the 
moduli fields in compactifications of type IIB string theory on a Calabi-Yau (CY) orientifold.\footnote{We consider only situations in which the warping is approximately constant over the compact manifold, except for one region, where a warped throat is generated (whose volume will be anyway smaller than the rest of the CY).}
In the simplest case of a single K\"ahler modulus $T$, the functions $W_0$ and $\rho$ do not depend on it at the perturbative level, due to holomorphy and the Peccei-Quinn symmetry $T\mapsto T+ic$. 
In the K\"ahler potential \eqref{EFTXVA}, the zeroth order term $K_0=-3\log(T+\bar{T})$ is known to be invariant (up to a K\"ahler transformation) under the full modular transformation $T\rightarrow (aT-ib)/(icT+d)$ (a generalisation of the shift symmetry). If  $X$ transforms appropriately, i.e. as a modular form of weight $\kappa$, the quadratic coeffcient is given by $K_2=\beta(T+\bar{T})^{-\kappa}$ (with $\beta$ a constant).
Moreover, if the linear term in $K$ is constant the only contribution of $X$ to the $F$-term scalar potential is the positive definite term
\be
V_{uplift}= e^KK^{-1}_{X\bar{X}}\left\| \frac{\partial W}{\partial X}\right\|^2=\frac{|\rho|^2}{\beta(T+\bar{T})^{3-\kappa}}\:.
\ee
This precisely coincides with the KKLMMT uplift term induced by an anti-D3-brane at the tip of a warped throat, if the modular weight is $\kappa=1$.\footnote{If $\kappa=0$ it would reproduce the original KKLT uplifting term but this holds only for the anti-D3-brane on an unwarped region and therefore the term is of order the string scale $V_{uplift}\sim M_s^4$ which, if included in the low energy EFT, would destabilise the vacuum by generating a runaway.
} Regarding the  gauge kinetic function,  even though a linear term in $X$ is allowed in general, the fact that the anti-D3-brane is localised at a particular point in the compactification manifold makes it difficult to have a direct coupling to gauge fields located at distant D3 or D7-branes. This then indicates that the anti-D3-brane in a warped throat can be described by an EFT with
\be
K=-3\log(T+\bar{T}) + \beta \frac{X\bar{X}}{T+\bar{T}}, \qquad W=W_0+\rho X, \qquad f=f_0,
\ee
where $c, \rho, W_0, f_0$ are constant and $|\rho|^2/\beta$ provides the warp factor in KKLMMT. 

Due to the nilpotency property of the superfield $X$, this K\"ahler potential can also be written in the form
\be\label{logKXXb}
K=-3\log\left(T+\bar{T}-\frac{\beta}{3} X\bar{X}\right)\:.
\ee
We then notice that in the regime when the EFT is valid, i.e. when the anti-D3-brane is at the tip of the throat, the $X$ superfield couples to $T$ in the K\"ahler potential in the same way as the superfield $\phi$ describing the D3-brane matter fields,\footnote{Here and in the following we will take a simplified model where we write down only one of the three complex superfields describing the D3-brane positions. Adding the other two would only complicate the expressions, without changing our results.} i.e. \cite{Grana:2003ek,Grimm:2004uq}
\be
K_{D3}=-3\log\left(T+\bar{T}-\frac{\alpha}{3}\phi\bar{\phi}\right) \sim -3\log\left(T+\bar{T}\right) + \alpha \frac{\phi\bar{\phi}}{T+\bar{T}} + \cdots \:,
\ee
where in the last step we have made an expansion in $1/(T+\bar{T})$. One could then conjecture that the only effect of $X$ in the K\"ahler potential is to shift the K\"ahler coordinate $T$ in the same way as the field $\phi$ does. We call this the {\it log hypothesis}, as it leads to write the $X$ inside the log as in \eqref{logKXXb}.

When we have both D3-branes and the anti-D3-brane in the background, generically we can write the K\"ahler potential as
\be\label{KparamEFT}
K=-3\log (T+\bar{T})+\alpha\frac{\phi\bar{\phi}}{(T+\bar{T})^\mu}+\beta\frac{X\bar{X}}{(T+\bar{T})^\kappa}+\gamma\frac{X\bar{X}\phi\bar{\phi}}{(T+\bar{T})^{\zeta}}+\cdots  
\ee
with modular weights $\mu=\kappa=1$ fitting the discussion above. Moreover, if $\phi$ and $X$ have modular weights $\mu$ and $\kappa$ respectively, the corresponding modular weight for the $\phi\bar{\phi} X\bar{X}$ term should be
$\zeta=\mu+\kappa$. In this case $\zeta=1+1=2$ \footnote{This can be seen to be consistent with the fact the $F^X$ contribution to the soft scalar masses is of order $m_{3/2}$ \cite{kqu}.}. 
This agrees with the {\it log hypothesis} introduced above, that would lead to the K\"ahler potential
\be
K_{no-scale}=-3\log\left(T+\bar{T}-\frac{\alpha}{3}\phi\bar{\phi}-\frac{\beta}{3}X\bar{X}\right)\label{Klog} \:.
\ee
In fact, expanding this in powers of $1/(T+\bar{T})$, one obtains \eqref{KparamEFT} with the only condition that $\gamma=\frac{\alpha\beta}{3}$. 
Notice that this is the standard no-scale form \cite{noscale}\,  of the K\"ahler potential.

In this paper we want to apply this EFT to the KKLT and LVS scenarios with matter living on D3-branes.
Notice that in KKLT the low energy effective theory is usually written in terms of the  fields with masses of order or below the gravitino mass. These include open string massless chiral fields as well as K\"ahler moduli. 
Supersymmetry is broken at the minimum of the scalar potential. Both the F-term of $X$ and the F-term of $T$ are different from zero (with $F_T\ll F_X$). Therefore, the full goldstino field would be  a combination of the fermion in $X$ and the fermion in $T$, with dominant $X$ component. 
In LVS already in the absence of the anti-brane the volume modulus $T_b$ breaks supersymmetry by having a non-vanishing F-term ($F_{T_b}\neq 0$). Including a nilpotent superfield in the effective action allows to consider the breaking of supersymmetry induced by fluxes and the one induced by the anti-brane on equal footing. Again the total goldstino will be a combination of the fermion components of $X$ and of the moduli. Even though the dominant component is usually the one from the $T_b$ field, for sequestered models the $X$ component is relevant and its contribution to the soft terms must be properly computed. We will address these issues in Sections \ref{Sec:KKLT} and \ref{Sec:LVS}.

\section{Warped flux compactifications and nilpotent fields}\label{Sec:WarpFlComp}

\subsection{Geometric approach}

We consider type IIB compactifications on Calabi-Yau (CY) orientifolds in presence of non-trivial background three-form fluxes. 
In the work by Giddings, Kachru and Polchinski (GKP) \cite{gkp} the solution of the ten dimensional (10D) equation of motion has been studied (see also \cite{Dasgupta:1999ss}). The  10D metric has the following form:
\be
ds_{10}^2=e^{2D} \eta_{\mu\nu} dx^\mu dx^\nu+e^{-2D} g_{mn} dy^m dy^n\label{eq:warpedmetric1} \:.
\ee
Here  $e^{2D(y)}\equiv h^{-1/2}(y) $ is the warp factor, with $h(y)$ satisfying a Poisson-like equation with sources coming from three-form fluxes and localised objects (brane/orientifold), and $g_{mn}$ is a  Calabi-Yau metric. For zero fluxes this function becomes a constant. For non-zero fluxes, it provides a factor in front of both the internal and external metric, that varies over the compact directions. As a result, the compact metric in no-longer CY (only conformally equivalent to it) and the 4D space-time metric is multiplied by the so-called {\it warp factor}. The warp factor acts as a redshift factor for the objects localised in the compact directions in regions where $e^{-2D}$ is large (like in \cite{Randall:1999ee}). In these regions, points that would be close in the unwarped CY metric are far away in the physical compact metric. These regions are called {\it warped throats} and their geometry is close to the Klebanov-Strassler (KS) throat \cite{ks}. The effective action of warped type IIB compactifications has been studied in \cite{DeWolfe:2002nn,deAlwis:2003sn,gm,Shiu:2008ry,Douglas:2008jx,Frey:2008xw,Martucci:2009sf,Chen:2009zi,Frey:2013bha,Martucci:2014ska}. Here we work in the approximation in which the warp factor is almost constant over the whole compact space, except for a single throat, whose volume is smaller than the rest of the space.

As pointed out in \cite{gm}, 
a constant shift of $e^{4D}$ leaves invariant the Poisson equation and can be identified with 
(a power of) the CY volume modulus. Furthermore a rescaling of the Calabi-Yau metric  $ds_{CY}^2$ to a unit-volume fiducial metric $ds_{CY_0}^2$ given by $ds_{CY}^2 = \lambda ds_{CY_0}^2$ can be compensated by a rescaling of the warp factor $e^{2D}= \lambda e^{2A}$. The warped metric can then be written schematically as
\be
ds^2_{10}= \vo^{1/3}\left(e^{-4A}+\vo^{2/3}\right)^{-1/2} ds^2_4+\left(e^{-4A}+\vo^{2/3}\right) ^{1/2}ds_{CY_0}^2\label{eq:warpedmetric2}\:,
\ee
which is equivalent to:
\be
ds^2_{10}= \left(1+\frac{e^{-4A}}{\vo^{2/3}}\right)^{-1/2} ds^2_4+\left(1+\frac{e^{-4A}}{\vo^{2/3}}\right)^{1/2}ds_{CY}^2\label{eq:warpedmetric3}\:.
\ee

Here $\Omega^2=\left(1+\frac{e^{-4A}}{\vo^{2/3}}\right)^{-1/2}$ is the redshift factor that, in a highly warped region defined by $e^{-4A}\gg \vo^{2/3}$,  behaves as $\Omega\sim e^A\vo^{1/6}\ll 1$.

Let us see the properties of this metric.
\begin{itemize}
\item{} In the large volume limit, i.e. $\vo^{2/3}\gg e^{-4A(y)}$ everywhere, the metric becomes the standard unwarped metric $ds_{10}^2=ds_4^2+\vo^{1/3}ds_{CY_0}^2=ds_4^2+ds_{CY}^2$.
\item{} In the largely warped regions, where $e^{-2A(y)}\gg \vo^{1/3}$, the internal part of the metric describing the warped throat becomes close to the KS geometry:
\be
ds^2_{10}= e^{2D_w(r)} ds^2_4+ e^{-2D_w(r)}\left( dr^2 + r^2 ds^2_{T^{1,1}}  \right)  \label{eq:warpedmetric4} \:,
\ee
where approximately $e^{-D_w(r)}\sim\frac{R}{r}$. This takes its maximal value at the tip of the throat ($r=r_0$): 
$e^{-D_w(r_0)} \sim \frac{R}{r_0}$, where $R$ is the typical size of the throat. In the GKP \cite{gkp} compactifications, $r_0$ measures the size of the three-sphere at the tip of the throat and is given by $r_0 \propto \vo^{1/6}  e^{-\frac{2\pi K}{3g_sM}} \ell_s$. Hence
\be
e^{4A_w(r_0)} \sim   e^{-\frac{8\pi K}{3g_sM}} \equiv e^{-4\varrho} \:,
\ee
where $g_s$ is the string coupling and $K$ and $M$ are the integral fluxes on the two dual three-cycles that define the throat.
 \item{} The warped volume $\vo_W$ that relates the 10D and 4D Planck masses is given by
 \be
 \vo_W=\int d^6 y \sqrt{g_{\small{CY}}} \, e^{-4D} = 
 \vo\int d^6 y \sqrt{g_{CY_0}}\left(1+\frac{e^{-4A}}{\vo^{2/3}}\right)\sim \vo\:,
 \ee
 where the last approximation is valid if the volume of the throat is small compared to the (large) volume of the CY. 
 
 \item{} The tension of an anti-D3-brane in the GKP background induces a positive term in the scalar potential. This term depends on the anti-D3 position $r_{{\rm D3}}$ in the compact space, i.e. whether it is in a warped or unwarped region:
 \ben
 2T_3\int d^4x\sqrt{-g_4}\sim 2M_s^4\frac{\vo^{2/3}}{e^{-4A(r_{{\rm D3}})}+\vo^{2/3}}&\sim &  \left\{ 
\begin{array}{lcl}
 \frac{e^{4A(r_{{\rm D3}})}}{\vo^{4/3}} & \rm{for} & e^{-4A(r_{{\rm D3}})}\gg \vo^{2/3} \\
 \frac{1}{\vo^2} & \rm{for} & \vo^{2/3}\gg e^{-4A(r_{{\rm D3}})} \\
\end{array} \right.
 \label{eq:uplift} \een
 where we are using $T_3=8\pi^3g_s\alpha'^2 \sim M_s^4\sim M_p^4/\vo^2$, with $M_s$ the string scale and $M_p$ the four dimensional Planck mass. Notice that the first expression gives  the uplifting term  ($2T_{\rm D3}$) in KKLMMT and the second one gives the one written in  KKLT. 
 
 \item{}
 In the presence of both large warping regions and large volume it is important to understand the conditions under which an effective field theory is valid. In these regions, we have 
 $e^{-4A}\gg\vo^{2/3}$. The massive string states of an anti-brane  sitting at the tip of such a throat are redshifted to lower masses and could be lighter than the gravitino mass $m_{3/2}
 \sim 1/\vo$ invalidating the use of a low energy effective field theory that neglects these states. Their mass is proportional the string scale $M_s\sim \vo^{-1/2}M_p$ redshifted by the factor $\Omega=\vo^{1/6}e^{A}$. Hence we need to require \cite{gm, bcdgmq, bcmq}:
\be
M_s^{w} \sim \Omega M_s\sim \frac{\vo^{1/6}e^{A}}{\vo^{1/2}}M_p=\frac{e^A}{\vo^{1/3}}M_p\gg m_{3/2}\sim \frac{W_0}{\vo}M_p
 \implies   e^{-A}\ll \vo^{2/3} \:.
\ee
Including the condition of being in a warp throat, the volume and the warp factor must satisfy $e^{-A}\ll \vo^{2/3}\ll e^{-4A} $.

\end{itemize}

\subsection{Brane/anti-brane dynamics\label{xxx}}

In this paper we mainly consider a visible sector realised by some D3-branes placed on top of a point-like singularity of the compact manifold. The susy breaking fluxes (together with perturbative and non-perturbative corrections to the effective action) induce soft terms on the worldvolumes of these branes that tend to stabilise the position of the D3-branes. It is a sensible question whether the attraction that the D3-branes feel towards the anti-D3-brane is enough to destabilise this minimum or can shift the D3-brane position away from the singularity (destroying the SM spectrum).

The potential generating such a force can be computed in the following way.
Geometrically, the D3-brane back-reacts on the geometry by modifying the harmonic function $h(r)$. Here we are assuming that the anisotropies of the internal directions are negligible (that for large volume of the compact manifold is plausible).
If the position of the D3-brane is $y_1$, 
the back-reaction of the D3-brane on the geometry induces a $y_1$ dependence on the warp factor:
\be
h(y,y_1)= h(y) + \delta h(y,y_1) \:.
\ee
At the tip of the throat (at $y=y_0$) the warp factor becomes
\be
e^{4D}\sim h(y_0)^{-1}\left(1- \frac{\delta h(y_0,y_1)}{h(y_0)}\right) \:.
\ee

Let us see how it works when the D3-brane is inside the throat. In the radial coordinates that are valid for the spherical symmetric (deformed) KS throat, the tip is at $r=r_0$ where the anti-D3-brane sits, while the D3-brane is at $r=r_1$. Moreover, we take
$r_0\ll r_1 \lesssim R$. In this case we know the approximate form of the metric and we can compute how the warp factor is modified. The radial position of the D3-brane is promoted to a scalar field, whose action is in this case
\be\label{potr1r0}
\int d^4x\sqrt{-g_4}\left[ \frac{1}{2} T_3 \partial_\mu r_{1}\partial_\mu r_{1} - 2T_3\frac{r_0^4}{R^4}\left(1-\frac{\ell_s^4}{R^4}\frac{r_0^4}{r_{1}^4}\right)\right]\:.
\ee
The last term gives the Coulombian attraction between the D3-brane and the anti-D3-brane.

We introduce the canonically normalised fields $\vec{\varphi}$,  that describe the position of the D3-brane in the six dimensional internal space.
Their relation to $r$ is 
\be
|\vec{\varphi}_1+\vec{\varphi}|=\sqrt{T_3} r_{1}\sim M_s^2 r_{1}
\ee
where we introduced the constant $|\vec{\varphi}_1|$
 to shift the origin of coordinates $\vec{\varphi}$. We consider $|\vec{\varphi}|\ll |\vec{\varphi}_1|$. 

Since $T_3\sim M_s^4\sim \vo^{-2}$  and $r_0/R\sim \vo^{1/6}e^{-\varrho}$ the scalar potential in units of $M_p$ can be written as
\begin{eqnarray}\label{eq:potphi}
V&=& \frac{e^{-4\varrho}}{\vo^{4/3}}\left(1-\frac{e^{-4\varrho}\vo^{2/3} }{\ell_s^4 |\vec{\varphi}_1+\vec{\varphi}|^4}
\right)\sim \frac{e^{-4\varrho}}{\vo^{4/3}}\left(1-\frac{e^{-4\varrho}\vo^{2/3}}{\ell_s^4|\vec{\varphi}_1|^4}\left(1-4\frac{\vec{\varphi} \cdot \vec{\varphi}_1}{|\vec{\varphi}_1|^2}+ 
10\left(\frac{\vec{\varphi} \cdot \vec{\varphi}_1}{|\vec{\varphi}_1|^2}\right)^2 +\cdots\right)\right) \:.\nonumber \\
&  &
\end{eqnarray}

When we move the D3-brane outside the throat, the potential \eqref{potr1r0} is still valid, with now $r_1$ being the distance between the D3-brane and the anti-D3-brane measured with the unwarped CY metric. If the D3-brane is at a generic point in the CY manifold, the distance from the anti-D3-brane is approximately $r_1 \sim \vo^{1/6} \ell_s$ and $|\vec{\varphi}_1|=r_1M_s^2=\vo^{1/6} \vo^{-1/2} M_p = \vo^{-1/3} M_p$ 
(with $\ell_s\sim \vo^{1/2}/M_p$). If we now plug these numbers into \eqref{eq:potphi} we obtain
\begin{eqnarray}\label{eq:potphi2}
V&\sim&  M_p^4\frac{e^{-4\varrho}}{\vo^{4/3}}\left(1-e^{-4\varrho}\left(1-4\vo^{1/3}\frac{|\vec{\varphi}|}{M_p}\cos\vartheta +
10\vo^{2/3}\frac{|\vec{\varphi}|^2}{M_p^2}\cos^2\vartheta +
\cdots\right)\right) \:,
\end{eqnarray}
where the angle $\vartheta$ measures the orientation of $\vec{\varphi}$ ($\cos\vartheta=\frac{\vec{\varphi}\cdot \vec{\varphi}_1}{|\vec{\varphi}_1|^2}$).

\begin{figure}[!ht]
\centering
\includegraphics[width=12.0cm]{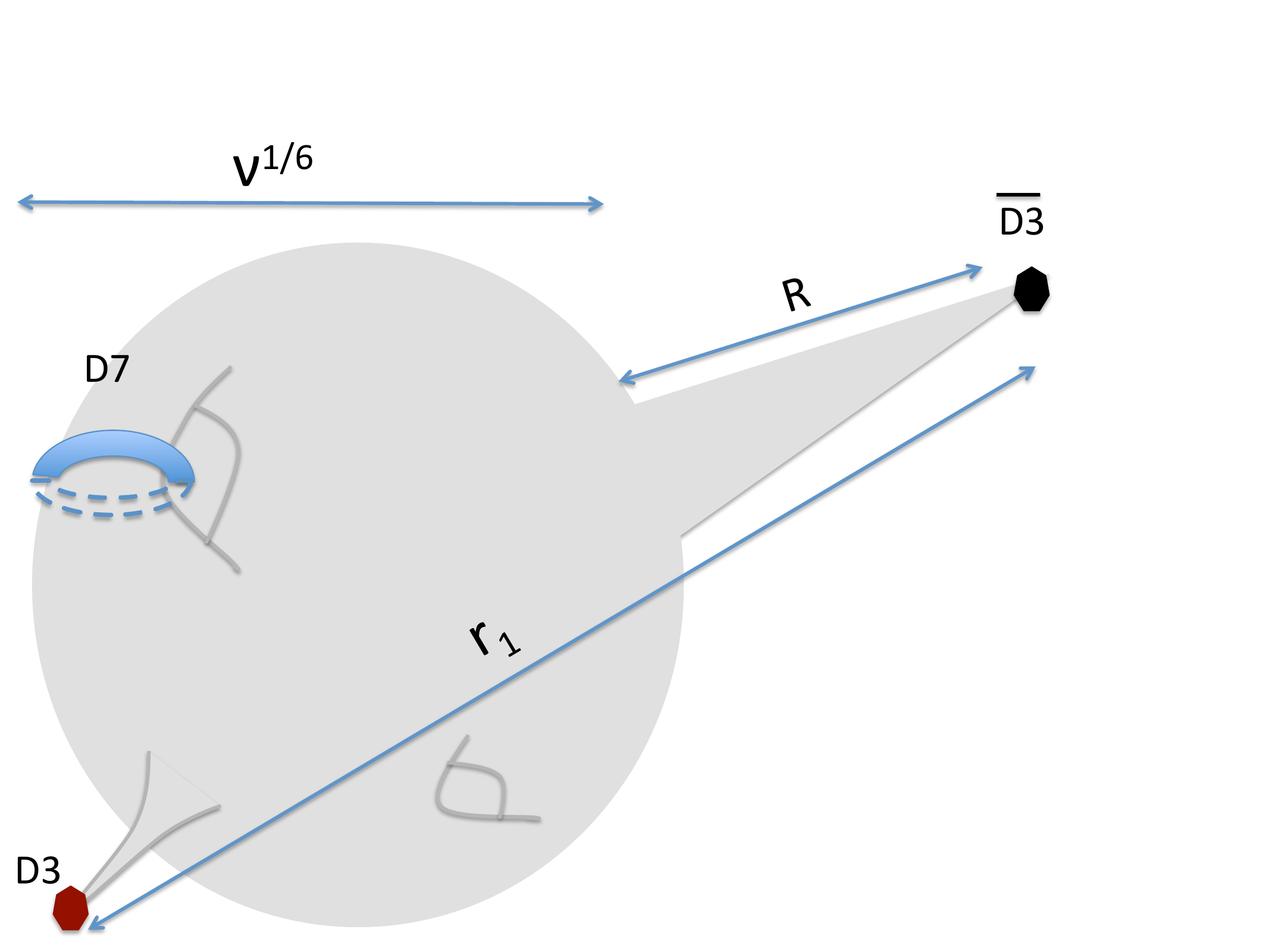}
\label{Fig1}
\caption{Cartoon description of the geometry and brane set-up}
\end{figure}

\subsubsection{Stability of D-branes at singularities: bounds on soft masses}

We can now consider the situation in which the D3-brane is at a singularity of the CY three-fold. At the singularity the D3-brane splits into a set of fractional branes with non-abelian gauge groups and chiral fermions. This can accommodate the visible MSSM sector. If the moduli are fixed in a non-supersymmetric vacuum, soft susy breaking terms are generated, giving a mass to the field $\hat{\varphi}$ (where we define $\hat{\varphi}\equiv |\vec{\varphi}|$) that stabilises it at zero. On the other hand, as we have just seen, the presence of an anti-D3-brane generates a Coulomb attraction for the D3-branes. If this is too strong, it can destabilise the location of the minimum. When this happens, the fractional D3-branes can recombine into a normal D3-brane that will start rolling towards the anti-D3-brane. As a result, the MSSM structure is destroyed. We now work out what are the bounds on the soft masses such that this does not happen.

 The $\hat{\varphi}$ dependent part of the potential is of order
 \be\label{potPhiOrder2}
 \delta V(\hat{\varphi}) = \frac{e^{-8\varrho}}{\vo} \hat{\varphi} M_p^3 +m_0^2 \hat{\varphi}^2 \:,
 \ee
 where the linear term comes from the Coulombian potential \eqref{eq:potphi} and where we
 have assumed that the soft term mass is dominant with respect to the quadratic negative part in \eqref{eq:potphi}, i.e.  $m_0^2 \gg \frac{e^{-8\varrho}}{\vo^{2/3}}M_p^2$.
The minimum of \eqref{potPhiOrder2} is at
 \be
 \hat{\varphi}\sim \frac{e^{-8\varrho}M_p^3}{2m_0^2\vo}\:.
 \ee
 Physically this non-zero vev for $\hat{\varphi}$ means that the D3-brane position is shifted from the original position by $\Delta r = \hat{\varphi}\ell_s^2$. 
 If this value were greater than the typical string length scale then it would mean that the presence of the anti-brane substantially affects the physics of the D3-brane system. Hence we need to impose $\Delta r \ll \ell_s$.
 
In order to have a de Sitter minimum, the uplifting term $e^{-4\varrho}/\vo^{4/3}$ has to be of the same order as $W_0^2/\vo^2$ in KKLT and as $1/\vo^3$ in LVS, that are the values of the potential evaluated on the AdS minimum (when the anti-D3-brane is not present). This implies that the warp factor must be of order respectively 
 $e^{-4\varrho}\sim W_0^2/\vo^{2/3}$ and $e^{-4\varrho}\sim 1/\vo^{5/3}$. When this happens, we have
\be
 \Delta r \sim \frac{e^{-8\varrho}M_p^3}{2m_0^2\vo}\ell_s^2 = \frac{M_p^2}{m_0^2}\frac{e^{-8\varrho}}{2\vo^{1/2}}\ell_s
= \left\{ \begin{array}{lcl}
 \frac{M_p^2}{m_0^2}\frac{W_0^4}{2\vo^{11/6}}\ell_s  && \mbox{for KKLT}\\   \\  
 \frac{M_p^2}{m_0^2}\frac{1}{2\vo^{23/6}}\ell_s && \mbox{for LVS} \\
 \end{array}\right. \:.
\ee 
Hence, $\Delta r \ll \ell_s$ if
\be
  \left\{ \begin{array}{lcl}
 \frac{m_0^2}{M_p^2} \gg \frac{W_0^4}{\vo^{11/6}}  &\qquad& \mbox{for KKLT}\\   \\  
 \frac{m_0^2}{M_p^2} \gg \frac{1}{\vo^{23/6}} &\qquad& \mbox{for LVS} \\
 \end{array}\right.  \:.
\ee  
Under these conditions, it is also valid that $m_0^2$ is leading with respect to the quadratic term in \eqref{eq:potphi}, that we had assumed at the beginning of this section ($ m_0 \gg \frac{W_0^2}{\vo^{4/3}} M_p\sim \frac{m_{3/2}^2}{M_{KK}}$ for KKLT and $ m_0\gg \frac{W_0^2}{\vo^{2}} M_p\sim \frac{m_{3/2}^2}{M_p}$ for LVS).

Notice that most models of supersymmetry breaking coming from KKLT and LVS satisfy the bounds. The only exception is the ultra-local case in LVS studied in references \cite{bckmq,ackmmq} for which the soft masses were precisely of order $m_0\sim m_{3/2}^2/M_p$ that is a borderline case.

\subsubsection{Supersymmetrising brane/anti-brane interactions}

We finish this section with an observation: by allowing the parameter $\rho$ in the superpotential \eqref{EFTXVA} to depend on the 
matter fields governing the D3-brane position, we are able to reproduce the Coulomb coupling \eqref{eq:potphi} 
from the effective supergravity point of view. 
The dependence of $\rho$ on $\hat{\varphi}$ should account for the modification of the anti-D3-brane contribution to
the potential due to the interaction with the D3-brane in the bulk.
Let us consider the simplest case of moduli stabilisation with all complex structure moduli and dilaton stabilised by fluxes and concentrate on the K\"ahler moduli and matter fields.
We study the effective field theory at low energies for one K\"ahler modulus $T$, with the volume determined by $\vo\sim (T+\bar{T})^{3/2}$,  one matter field $\phi$ representing the position of a D3-brane and the anti-D3-brane superfield $X$. The field $\phi$ is the proper K\"ahler coordinate and it is related to the field $\hat{\varphi}$ that we used before:
\be\label{phivarphirel}
 \hat{\varphi}=\frac{|\phi|}{\sqrt{3(T+\bar{T})}}\sim \frac{|\phi|}{\vo^{1/3}}  \:.
 \ee
For this analysis, we take the K\"ahler potential \eqref{KparamEFT}
and the superpotential
\be
W=W_0(U,S)+ W_{np}(U,S,T)+\rho(U,S,\phi) X \:,
\ee
where we allow $\rho$ to depend on $\phi$. 
The contribution of the $X$ superfield to the scalar potential, given by the K\"ahler potential and the superpotential just presented, is very simple to extract:
\be
V_{F_X}=e^K K^{-1}_{X\bar{X}} \left\|D_XW\right\|^2=\vo^{-2} \vo^{2/3}\beta^{-1}\left\|\frac{\partial W}{\partial X}\right\|^2=\frac{|\rho|^2}{\beta\vo^{4/3}}\sim\frac{|\rho|^2}{\beta\left(T+\bar{T}\right)^{2}} \:.
\label{eq:potsg}
 \ee
 
We now consider the $\phi$ dependence of $\rho$, by expanding it around $\phi=0$:
 \be\label{rhoExpPhi}
 \rho=\rho_0+\delta\rho=\rho_0+\rho_1 \phi+ 
 \cdots
 \ee
 where $\rho_0$ gives the constant KKLMMT uplift term if $|\rho_0|^2 \sim e^{-4\varrho}$.  We now plug \eqref{rhoExpPhi} into \eqref{eq:potsg} and we expand this around $\phi=0$:
\ben
 V_{F_X} &\sim &  \frac{1}{\beta\vo^{4/3}}\left(|\rho_0|^2+2{\rm Re}(\rho_0^*\rho_1 \phi)+  \cdots\right) \\ 
 &\sim &   \frac{1}{\beta\vo^{4/3}}\left(|\rho_0|^2+2\vo^{1/3}{\rm Re}(\rho_0^*\rho_1 \hat{\varphi}e^{i\tilde\vartheta})+ \cdots\right)\:,\nonumber 
 \een
where $\tilde{\vartheta}$ is the phase of $\phi$ and we used \eqref{phivarphirel} to substitute $|\phi|=\vo^{1/3}\hat\varphi $. 
 
 We can now compare this expression with the analogous expansion \eqref{eq:potphi2} of the Coulomb potential.
 We realise that the volume dependence exactly matches and that
 \be
 |\rho_0|\sim e^{-2\varrho} \qquad\mbox{and} \qquad  |\rho_1|\sim e^{-6\varrho} \:.
 \ee
For real $\rho_i$, we also have that the phase $\tilde \vartheta$ of $\phi$ matches with 
the angle $\vartheta$ between $\vec{\varphi}$ and $\vec{\varphi}_1$.
 
 This analysis suggests that the interaction between the anti-D3-brane at the tip of the throat and the D3-brane in the bulk
 can be reproduced at the level of the supergravity EFT by letting the parameter $\rho$ in the superpotential \eqref{EFTXVA} depend on the D3-brane position moduli 
 $\phi$.

We end this section with a curiosity.
 If we generalised the K\"ahler potential \eqref{KparamEFT} by taking a different
factor in front of the $X\bar{X}$ term, i.e. instead of $\frac{\beta}{\vo^{2/3}}$ we took 
$\frac{\beta}{\vo^{2/3}} + b(U,\bar{U})$, then 
the uplift term $e^{K}K^{-1}_{X\bar{X}} |\rho_0|^2$ could be written as:
\be
V_{up}=e^{K}K^{-1}_{X\bar{X}} |\rho_0|^2=\frac{|\rho_0|^2}{\vo^2}\frac{\vo^{2/3}}{\beta+b\vo^{2/3}}\label{eq:uplift2}
\ee
Notice that for $b\sim e^{4A}$, the equation \eqref{eq:uplift2} would reproduce exactly the 
general result of \eqref{eq:uplift} interpolating between KKLT and KKLMMT uplift. In particular, if 
$b\ll \beta\vo^{-2/3}$ then we would recover the warped KKLMMT uplift, while if the volume dominated over the warp factor,
$b\vo^{2/3}\gg \beta$ then we would recover the unwarped uplifting originally proposed in KKLT.

\section{Nilpotent goldstino in KKLT} \label{Sec:KKLT}

In this section we deform the EFT of the type IIB KKLT flux vacua, by introducing the nilpontent chiral 
superfield $X$. As we have discussed, this produces an uplift term and breaks supersymmetry. 
By using standard supergravity techniques we will compute the soft terms that are generated in the visible sector realised on D3-branes at singularities.

\subsection{Scalar potential}

 Following the standard KKLT moduli stabilisation procedure, we assume that the dilaton and the complex structure moduli have been fixed 
 at high scale and hence are integrated out form the EFT.
The K\"ahler potential for the remaining fields, i.e. the K\"ahler modulus $T$ (whose real part controls the CY volume: $\vo=\tau^{3/2}$, with $\tau=\,$Re$T$), the matter field $\phi$ and the nilpotent field $X$, is\footnote{The blow up modulus $\tau_{D3}$ is fixed to zero by the higher order D-term potential and it is then integrated out in the EFT we are considering here \cite{Cicoli:2012vw}.}
\begin{equation}
 K = -3\log(T+\bar{T}) + \tilde{K}_i\ \phi \bar{\phi} + \tilde{Z}_i\ X \bar{X} + \tilde{H}_i\ \phi \bar{\phi} \ X \bar{X} + ...
\label{paramK\"ahler}
 \end{equation}
where  $\tilde{K}_i$ and $\tilde{Z}_i$ are the matter metric for the matter fields living on D3-branes and the nilpotent goldstino respectively and $\tilde{H}_i $ is the quartic interaction between the matter fields and the nilpotent goldstino. Following the discussion in Section \ref{Sec:NilpGold}, these are given by
\begin{equation}
 \tilde{K}_i  = \frac{\alpha}{\tau}\:, \qquad \tilde{Z}_i  = \frac{\beta}{\tau}\:,\qquad \tilde{H}_i  = \frac{\gamma}{\tau^{2}}\:,
\label{mattermetric}
 \end{equation}
where the scaling of $\tilde{K}_i$ with $\tau$ is due to the modular weight of the matter fields on D3-branes \cite{Conlon:2006tj,Aparicio:2008wh}. The superpotential is
\begin{equation}
 W \ =\ W_{0} +\rho X  + A\ e^{-a T}\:,
\label{Wnil}
 \end{equation}
 where we included the non-perturbative contribution necessary to stabilise the K\"ahler modulus in KKLT.
 The supergravity F-term scalar potential is determined by $K$ and $W$ (respectively (\ref{paramK\"ahler}) and (\ref{Wnil})). Here and in the following we are measuring everything in units of the 4D Planck mass, i.e. we take $M_p=1$. 
 
The supergravity potential is determined in terms of the K\"ahler potential and the superpotential by the formula
 \begin{equation}\label{sugrapotential}
  V = e^{K} \left(  K^{I\bar{J}} D_I W D_{\bar{J}}\bar{W} - 3 |W|^2  \right) \:, \qquad \mbox{with} \qquad D_IW = \partial_I W + K_I W \:.
 \end{equation}
In this equation,  $I,J$ run over the chiral superfields $T,\phi,X$,  $K^{I\bar{J}}$ is the inverse of the matrix $K_{I\bar{J}}\equiv \partial_I\partial_{\bar{J}} K$ and $K_I\equiv \partial_IK$. Sometimes we write this formula as
\begin{equation}\label{sugrapotential2}
V =  F^{I} F_I  - 3 m_{3/2}^2
\end{equation}
where $F_I \equiv e^{K/2} D_IW$ and $F^I \equiv e^{K/2}K^{I\bar{J}} D_{\bar{J}}\bar{W}$ are the F-term controlling supersymmetry breaking, and $m_{3/2}\equiv e^{K/2}|W|$ is the gravitino mass.

By plugging the expressions \eqref{paramK\"ahler} and \eqref{Wnil} into \eqref{sugrapotential},  after some manipulations one obtains 
\begin{equation}\label{ScPotKKLTUp}
 V = \left(V_{KKLT} + V_{up}\right) + \frac{2}{3}\left(  (V_{KKLT}  +   V_{up}) + \frac{1}{2} V_{up} \left(1-\frac{3 \gamma}{\alpha \beta} \right)  \right) \ |\hat{\phi}|^2\:,
\end{equation}
where  $\hat{\phi}$ is the canonically normalised matter scalar field and $V_{KKLT}$ is the standard KKLT potential (without the uplifiting term):
\begin{equation}
 V_{KKLT} \ =\ \frac{2\ e^{-2 a \tau} a A^2}{s\ \mathcal{V}^{4/3}} + \frac{2\ e^{-2 a \tau} a^2 A^2}{3s\ \mathcal{V}^{2/3}}\ - \frac{2\ e^{- a \tau} a A\ W_0}{s\ \mathcal{V}^{4/3}}\:,
\label{kklt3} 
 \end{equation}
where $s=1/g_s$ is the real part of the axiodilaton $S=e^{-\phi}-iC_0$ that is fixed at higher scale by three-form fluxes (hence at this level of the EFT $s$ is just a parameter).\footnote{
Here and in the following we are neglecting the $e^{K_{cs}}$ factor in front of the scalar potential, where $K_{cs}$ is the K\"ahler potential for the complex structure moduli. Since they are fixed at higher scales, this is just a flux dependent parameter in the studied EFT. This factor would appear in front of all the relevant scales, like the gravitino mass and the soft masses, but is not affecting our results, as we are giving the soft masses in terms of the gravitino one.}
Moreover, recall that  $\tau$ is the real part of the K\"ahler modulus $T = \tau + i \psi$. The imaginary part of this modulus behaves like an axion and develops a minimum for  $\psi = \pi/a$. 
This value is responsible for the minus sign in the third term in  (\ref{kklt3}).

The uplift term $V_{up}$ coming from $F^XF^*_X$ is
\begin{equation}\label{VupKKLT}
 V_{up} = \frac{\rho^2}{2\beta s \tau^{2}} \:.
\end{equation}
Minimising the scalar potential, one finds that at the  minimum
\begin{equation}
  W_0 = e^{-a \tau} A \left(1 + \frac{2}{3} a \tau + \frac{e^{2 a \tau} \rho^2}{2 \beta a^2 A^2 \tau}\right) \:.
  \label{kkltmin4}
\end{equation}
Plugging this condition into the scalar potential \eqref{ScPotKKLTUp}, one obtains its value at the minimum, that is
\begin{equation}\label{ScPotKKLTtot}
 V = \left(V_0^{KKLT} + V_{up}\right) + \frac{2}{3}\left(  (V_0^{KKLT}  +   V_{up}) + \frac{1}{2} V_{up} \left(1-\frac{3 \gamma}{\alpha \beta} \right)  \right) \ |\hat{\phi}|^2 \:,
\end{equation}
where
\begin{equation}
 V_0^{KKLT} = - \frac{2 e^{-2 a \tau} a^2 A^2}{3s\ \tau}  = -\frac{3\ W_0^2}{2s \mathcal{V}^{2}} = -3 m_{3/2}^2<0\:.
\label{kkltCC}
 \end{equation}

Notice that without the uplift term ($V_{up}\equiv 0$), the minimum would be supersymmetric, i.e. we would have $D_TW=0$, the cosmological constant would be negative and the squared masses would be tachyonic (this does not signal an instability, as we have a supersymmetric AdS vacuum). 
After adding the uplift term, the minimum is no more supersymmetric; in particular, using the minimum condition  (\ref{kkltmin4}) one has 
\begin{equation}
 D_T W = -\frac{1}{4} \frac{e^{a \tau} \rho^2}{a^2 A\ \tau^2}\:.
\end{equation}

The flux dependent parameter $\rho$ can be tuned to make the cosmological constant zero or extremely small (positive). This happens when the uplift term \eqref{VupKKLT} is (approximately) equal to the KKLT contribution \eqref{kkltCC}, i.e. when
 \begin{equation}\label{dScondRho}
  \rho^2 = \frac{4 \beta}{3}\ \tau \ e^{-2a\tau} a^2 A^2 \:.
 \end{equation}
After imposing the null cosmological constant  condition \eqref{dScondRho}, the KKLT minimum condition (\ref{kkltmin4}) gives us 
\begin{equation}
 W_0 = \frac{e^{-a \tau} A}{3}\ \left(2a \tau + 5\right) \:.
 \label{kkltmin3}
 \end{equation}

The scalar masses can be read off from the scalar potential \eqref{ScPotKKLTtot} evaluated at the minimum. After tuning $\rho$ as in \eqref{dScondRho}, one obtains (remember we have set $M_p=1$)
\begin{equation}
 m_0^2 = \frac{1}{3} V_{up} \left(1-\frac{3 \gamma}{\alpha \beta} \right) =
\left(1-\frac{3 \gamma}{\alpha \beta} \right)  \frac{2 a^2 A^2 e^{-2a\tau} }{9 s \tau}\:.
\label{scalarparam}
\end{equation}
From (\ref{scalarparam}) we can see that there are two different terms that compete: the first one comes form the contribution of the nilpotent field $X$, while the second one, with opposite sign, comes from the quartic interaction term in \eqref{paramK\"ahler} between the matter fields and $X$. Depending on the values of $\alpha,\beta,\gamma$ these masses might be tachyonic.

Interestingly, notice that if the {\it log hypothesis} is valid, i.e. the K\"ahler potential takes the form \eqref{Klog}, there is a cancellation inside the parenthesis in \eqref{scalarparam} that makes the scalars massless ($m=0$).  In fact the {\it log hypothesis} is realised when $\gamma=\frac{\alpha\beta}{3}$.
In this case, the contribution of the susy breaking term to scalar masses coming from the KKLT K\"ahler potential is exactly zero (or as small as the cosmological constant value in our dS universe). Hence the subleading contributions to the scalar masses, at next order of approximation, would be the  dominant one. In particular the (always present) contribution coming from anomaly mediation, which is negative for sleptons, may dominate (see the discussion in Appendix \ref{Sec:AnomalyMed}). This would imply that the pure KKLT is unstable, since the anomaly mediated contributions  produce always tachyonic scalar masses in a dS vacuum. Therefore non-vanishing contributions should also be considered at the next order in the approximation.

\subsection{$\alpha'$ corrections to KKLT}\label{KKLTalphaCorr}
In this section we are going to study the $\alpha'$ contributions to the KKLT dS minimum. For that purpose we are going to use again an effective field theory, where now the form of the K\"ahler potential is modified:
\begin{equation}
  K = -2\log \left(\tau^{3/2}- \hat\xi \right)  + \tilde{K}_i\ \phi \bar{\phi} + \tilde{Z}_i\ X \bar{X} + \tilde{H}_i\ \phi \bar{\phi} \ X \bar{X} + ...
\label{paramK\"ahleralpha}
  \end{equation}
where $\hat\xi = s^{3/2} \xi /2$ and $\xi$ is a constant of order one depending on the Euler characteristic of the Calabi-Yau manifold $\chi$ \cite{bbhl}.\footnote{In the most conservative view, the constant $\xi$ is given by \cite{bbhl}  as $\xi= -\frac{\zeta(3)\chi}{2(2\pi)^3}$. Recently, \cite{SavelliMinasian} found that the presence of an orientifold O7-plane wrapping the divisor $D$ modifies $\xi$ by shifting the Euler characteristic as $\chi\mapsto \chi + 2\int_{CY} D^3$
} The matter field metric and the quartic coupling defined in (\ref{mattermetric}) can also receive  $\alpha'$ corrections that can be parametrised in the following way:
\begin{equation}
 \tilde{K}_i = \frac{\alpha_0}{\mathcal{V}^{2/3}}\left(1-\alpha_1 \frac{\xi s^{3/2}}{\mathcal{V}} \right),
\label{mattermetric2}
 \,\,\,\,
 \tilde{Z}_i = \frac{\beta_0}{\mathcal{V}^{2/3}}\left(1-\beta_1 \frac{\xi s^{3/2}}{\mathcal{V}} \right),
 \,\,\,\,
 \tilde{H}_i = \frac{\gamma_0}{\mathcal{V}^{4/3}}\left(1-\gamma_1 \frac{\xi s^{3/2}}{\mathcal{V}} \right) \:.
\end{equation}
One can recover the theory from the logarithmic K\"ahler potential
\begin{equation}
  K \ =\ -2\log\left(\left(T+\bar{T}- \frac{\alpha}{3} \phi \bar{\phi} - \frac{\beta}{3} X\bar X\right)^{3/2} - \hat\xi \right) 
\end{equation}
if the parameters in  (\ref{paramK\"ahleralpha}) satisfy the relations
\begin{equation}
 \gamma_0\ = \frac{\alpha_0\beta_0}{3}\,, \qquad
 \alpha_1 \ = \beta_1\ = 1/2\,, \qquad
 \gamma_1 = 5/4 \:.\label{log2}
\end{equation}

The scalar potential for a theory with the K\"ahler potential \eqref{paramK\"ahleralpha} and the superpotential \eqref{Wnil} is given by
\begin{equation}
  V = \left(V_{\alpha'KKLT} + V_{\alpha'up}\right) + \left(  \frac{2}{3}(V_{\alpha'KKLT} + V_{\alpha'up})  + \Theta_{up} + \Theta_{\alpha'} + \Theta_{\alpha'/up}    \right) |\hat{\phi}|^2
  \label{kklt4}
\end{equation}
where by $ V_{\alpha'KKLT}$ we mean the standard KKLT potential $V_{KKLT}$ (given in (\ref{kklt3})) corrected by the $\alpha'$-contributions:
\begin{equation}
 V_{\alpha'KKLT} = V_{KKLT} + \frac{e^{-2a\tau}\, \sqrt{s}\, \xi \, a^2 A^2}{6\, \mathcal{V}^{5/3}} + \frac{e^{-a\tau}\,  \sqrt{s}\, \xi\, a A\, W_0}{2\, \mathcal{V}^{7/3}} + \frac{3\,  \sqrt{s}\, \xi\, W_0^2}{8\, \mathcal{V}^{3}} \:.
\end{equation}
The $\alpha'$ correction to the $V_{up}$ is given by
\begin{equation}
 V_{\alpha' up} = \frac{\rho^2}{2\beta_0\, s\, \mathcal{V}^{4/3}} \left( 1- \frac{s^{3/2} \xi}{\mathcal{V}} (1-\beta_1) \right)
\end{equation}
and the contributions  $\Theta_{up}$, $\Theta_{\alpha'}$ and $\Theta_{\alpha'/up}$ to the scalar masses correspond to the pure uplifting, the pure $\alpha'$ and the mixed uplifting-$\alpha'$ respectively. Similarly to the case studied in the previous section,
\begin{equation}
 \Theta_{up} = \frac{1}{3}  V_{\alpha' up} \left(1-\frac{3 \gamma_0}{\alpha_0 \beta_0} \right)\:,
 \label{mass1}
\end{equation}
which is exactly zero for the logarithmic K\"ahler potential according to (\ref{log2}). The pure $\alpha'$ contribution to the mass is given by
\begin{equation}
 \Theta_{\alpha'} = \frac{5 \sqrt{s}\, \xi}{2} \left( \frac{e^{-2a\tau} \, a^2 A^2}{9\, \mathcal{V}^{5/3}} -  \frac{e^{-a\tau}\, a A\, W_0}{3\, \mathcal{V}^{7/3}} + \frac {W_0^2}{4\, \mathcal{V}^{3}}   \right)(3\alpha_1 -1)\:.
 \label{mass2}
\end{equation}
The contribution coming from a combined effect of the nilpotent goldstino and the $\alpha'$ corrections is given by
\begin{equation}
  \Theta_{\alpha'/up}  = \frac{\sqrt{s}\, \xi\, \rho^2}{2\, \beta_0\, \mathcal{V}^{7/3}}\, \frac{\gamma_0}{\alpha_0 \beta_0}\, (\gamma_1 - \alpha_1 -\beta_1)\:.
  \label{mass3}
\end{equation}

We minimise the scalar potential (\ref{kklt4}) and restrict $\rho$ in order to have (approximately) null cosmological constant. This leads to a condition like (\ref{kkltmin3}) that is now (at leading order in $1/\tau$ expansion)
\begin{equation}
  W_0 =  e^{-a \tau} A\ \left(\frac{5}{3} + \frac{2}{3} a\ \mathcal{V}^{2/3} + \frac{a}{3} \frac{s^{3/2} \, \xi}{\mathcal{V}^{1/3} } \right)\:.
\end{equation}
In the new dS non-supersymmetric minimum there is a hierarchy between the contributions to the scalar mass. The biggest will be the one coming from the pure uplifting effect
\begin{equation}
  \Theta_{up} = \frac{2}{3}\, \frac{e^{-2a \tau} a^2 A^2}{s\, \mathcal{V}^{2/3}}     \left(1-\frac{3 \gamma_0}{\alpha_0 \beta_0} \right) 
  \sim \frac{W_0^2}{\mathcal{V}^{2}}\:.
\end{equation}
Notice however that this term is zero if  $\gamma_0 =\frac{\alpha_0\beta_0}{3}$. The second relevant term is
\begin{equation}
  \Theta_{\alpha'/up}  =\frac{2}{3} \frac{e^{-2a \tau} \sqrt{s}\, \xi\,  a^2 A^2}{ \mathcal{V}^{5/3}}\, \frac{\gamma_0}{\alpha_0 \beta_0}\, (\gamma_1 - \alpha_1 -\beta_1) \sim \frac{W_0^2}{\mathcal{V}^{3}}
\end{equation}
and finally the pure $\alpha'$ corrections  are very suppressed due to a cancellation at leading order in this minimum:
\begin{equation}
 \Theta_{\alpha'} =\frac{125}{72}  \frac{e^{-2a\tau}\sqrt{s}\, \xi \, A^2}{\mathcal{V}^{3}}  (3\alpha_1 -1) \sim \frac{W_0^2}{\mathcal{V}^{13/3}}\:.
\end{equation}
Notice that this $\alpha'$ correction is also proportional to the non-perturbative effect. Hence, sending this contribution to zero would make $\Theta_{\alpha'} $ vanish. This is the result of the fact that in KKLT susy is broken by the non-perturbative effects, not by the $\alpha'$ corrections. They have an impact in the soft terms but they are negligible in determining the minimum  as it has just been discussed.

We see that in the  KKLT scenario with $\alpha'$ corrections the masses are not zero even when we have a log form of the K\"ahler potential (like in \eqref{paramK\"ahleralpha}). This can make the scenario stable against the anomaly mediated contributions or the Coulombian attraction.

\subsection{Soft terms and F-terms in $\alpha'$KKLT with nilpotent goldstino}

\subsubsection{F-terms}\label{FtermsKKLT}
The susy breaking is determined by the F-term, that in supergravity are given by $F^I \equiv e^{K/2}K^{I\bar{J}} D_{\bar{J}}\bar{W}$. In the case under study, the dominant effect comes from the anti-D3-brane nilpotent superfield. Its F-term is
\begin{equation}
 F^X = \sqrt{\frac{3}{\beta_0}} \mathcal{V}^{1/3} \, m_{3/2}\:,
\end{equation}
where we remind that $m_{3/2}\equiv e^{K/2}|W|$ is the gravitino mass.
The presence of the nilpotent superfield is also inducing an F-term for the K\"ahler modulus:
\begin{equation}
 F^T = -\frac{2}{a} \left(1+ \frac{1}{a\,  \mathcal{V}^{2/3}} \right) \, m_{3/2} \:.
\end{equation}

Now we want to compute $F^S$. This takes contributions from $D_SW$ and, because of the mixing induced by $\alpha'$ corrections, fom $D_TW$.
At leading order in our approximation $D_SW=0$, as it appears squared in the leading term of the scalar potential. 
When we include non-perturbative and $\alpha'$ corrections and we consider the uplift term, these induce corrections to $D_SW$ in the non-supersymmetric minimum. In order for our expansion to work, these can at most induce a $D_SW$ of the order of the non-perturbative correction to $W$. More precisely 
$D_S W \sim \frac{e^{-a \tau} A}{2s}$,
since at leading order $D_SW_{\rm flux}=0$. Of course this is just an upper bound. To obtain the right value of $D_SW$ one should minimize the full potential. It might well be that $D_SW$ is much smaller than the above estimation. For this reason, we write its value as\footnote{Notice that if $D_SW\sim \frac{e^{-a \tau} A}{2s}$, its contribution to the scalar potential is subleading in the $1/\vo$ expansion with respect to the KKLT potential (even considering $\alpha'$ corrections).} 
\begin{equation}
 D_S W \sim \frac{e^{-a \tau} A}{2s} \, \omega_s   \:,
\label{dilatonmin}
\end{equation}
where $\omega_s\lesssim 1$ parametrises our ignorance and it takes values at most of order one.
If $\omega_s$ is sufficiently large, the F-term $F^{S}$ is dominated by the $F_S$ contribution such that
\begin{equation}
F^S = \frac{3s\, \omega_s}{a\, \mathcal{V}^{2/3}} 
m_{3/2}\:.
\end{equation}
However, if $\omega_s$ is sufficiently small, the $F_T$ contribution to $F^{S}$ dominates and in this case
\begin{equation}
F^S =\frac{9}{2} \frac{s^{5/2}\xi}{a\,\vo^{5/3}} m_{3/2} \:.
\end{equation}

\subsubsection{Soft terms}
In this section, expanding on the work of \cite{kqu}, we are going to write the soft-terms as functions of the gravitino mass for a theory with the K\"ahler potential (\ref{paramK\"ahleralpha}). 
In a supersymmetric effective field theory one can use the general expressions for soft terms \cite{soft}:
\ben\label{softGenExpr}
m_0^2 & = & V_0+m_{3/2}^2 - F^I\bar{F}^J\partial_I\partial_{\bar{J}} \log \hat{K} \,, \nonumber \\ 
M_{1/2}&=& \frac{1}{f+f^*}F^I\partial_I f\,,  \\
A_{ijk} & = & F^IK_I+F^I\partial_I\log Y_{ijk}-F^I\partial_I\log\left(\hat{K}_i \hat{K}_j \hat{K}_k\right) \,.\nonumber
\een
Here, indices $i,j,k$ label different matter fields, indices $I,J$ run over moduli fields and the $X$ field. $f$ is the holomorphic gauge kinetic function of the visible sector, depending only on the moduli fields and the dilaton, $\hat{K}$ is the matter K\"ahler metric (including the $X$ dependence, i.e. $\hat{K}_i=\tilde{K}_i+\tilde{H}_i X\bar{X}$, with $\tilde{K}_i$ and $\tilde{H}_i$ given in \eqref{mattermetric2}) and $Y_{ijk}$ are the Yukawa couplings among matter fields.

The scalar masses have been already given in \eqref{scalarparam}, as they can be read directly from the scalar potential \eqref{ScPotKKLTtot}. At leading order they can be written as
\begin{equation}\label{KKLTsofscal}
 m_0^2 = \left(1-\frac{3 \gamma_0}{\alpha_0 \beta_0} \right) m_{3/2}^2 + \frac{s^{3/2} \, \xi}{\mathcal{V}} \, \frac{3 \gamma_0}{\alpha_0 \beta_0} \, (\gamma_1 - \alpha_1 -\beta_1)\, m_{3/2}^2 \:.
\end{equation}
This result agrees of course with the derivation via \eqref{softGenExpr}.
Notice that in general the first term in \eqref{KKLTsofscal} is the dominant one and depending on the values of $\alpha_0,\beta_0, \gamma_0$ the square masses may be positive or negative. However when the K\"ahler potential takes the log structure \eqref{Klog}, i.e. when (\ref{log2}) are fulfilled, this term vanishes and the scalar masses become
\begin{equation}
 m_0^2 =  \frac{s^{3/2} \, \xi}{4\, \mathcal{V}} \, m_{3/2}^2\:,
\end{equation}
which are positive definite and therefore non-tachyonic. Notice also the suppression with respect to the gravitino mass.

The masses of the gauginos will depend on the form of the gauge kinetic function. For the D3-branes in KKLT  $f=S  $, therefore the gaugino masses will be dominated by the $F^S$ term. For the case of $\omega_s \sim \mathcal{O}(1)$ 
\begin{equation}
 M_{1/2} =    \pm\frac{3}{2a\, \mathcal{V}^{2/3}} 
 m_{3/2}\:,
\end{equation}
where the relative sign $\pm$ refers to the choice of $W_0\gtrless 0$. The trilinears are given by
\begin{equation}
 A_{ijk} = \frac{-3}{2a\, \mathcal{V}^{2/3}} \left(  1-  \frac{s^{3/2} \, \xi}{\mathcal{V}^{1/3}} \right) \left(1-s \partial_s \log(Y_{ijk}^{(0)})\right) \, m_{3/2}\:,
\end{equation}
where $ Y_{ijk}^{(0)}$ are the holomorphic Yukawa couplings. Hence the relation between trilinears and gauginos is:
\begin{equation}
A_{ijk} = -  \left(1-s \partial_s \log(Y_{ijk}^{(0)})\right) M_{1/2}\:.
\end{equation}

In the other limit, where $\omega_s$ is so small that the $D_TW$ contribution dominates,  the gaugino masses are generated at the $\alpha'$ level:
\begin{equation}
 M_{1/2} = \frac{9}{4}\frac{s^{3/2}\xi}{a \mathcal{V}^{5/3}}\, m_{3/2}\:.
\end{equation}
In this case the trilinears will be also modified, and could be written in terms of gaugino masses as
\begin{equation}
A_{ijk} = -  \left(\frac{5}{3}-2\alpha_1-s \partial_s \log(Y_{ijk}^{(0)})\right) M_{1/2}\:.
\end{equation}

\section{Nilpotent goldstino in LVS}\label{Sec:LVS}

\subsection{Vacuum structure}

In this section we will repeat the previous analysis in the Large Volume Scenario (LVS) \cite{Balasubramanian:2005zx}. We will study how the explicit antibrane uplift contribution to the potetial affects the soft terms. In fact, as reported in the general discussion started in
\cite{bckmq, ackmmq}, the uplift mechanism is relevant for sequestered models, in particular for branes at singularities.

Following the same notation as before, the K\"ahler potential can be written as 
\begin{equation}
K = -2\log \left(\mathcal{V}- \hat\xi \right)  
+ \tilde{K}_i\ \phi \bar{\phi} + \tilde{Z}_i\ X \bar{X} + \tilde{H}_i\ \phi \bar{\phi} \ X \bar{X} + ... \:.
\label{K\"ahlerlvs}
\end{equation}
We will concentrate on the simplest and most representative LVS example, with two  K\"ahler moduli $T_s$ and $T_b$ with real parts  $\tau_b$ and $\tau_s$ that determines the CY volume $\mathcal{V} = \tau_b^{3/2}-\tau_s^{3/2}$. The coefficients $\tilde{K}_i$, $\tilde{Z}_i$ and $\tilde{H}_i$ are defined as in \eqref{mattermetric2}. 
The superpotential is given by
\begin{equation}
 W \ =\ W_{flux} +\rho X  + A\ e^{-a_s T_s}.
\end{equation}
For such a theory the supergravity scalar potential will take the general form
\begin{equation}
  V = \left(V_{LVS} + V_{\alpha'up}\right) + \left(  \frac{2}{3}(V_{LVS} + V_{\alpha'up})  + \Theta_{up} + \Theta_{\alpha'} + \Theta_{\alpha'/up}    \right) |\hat{\phi}|^2 \:,
  \label{Vtotlvs}
\end{equation}
where $V_{LVS}$ is the standard LVS potential\footnote{Interestingly, it can also be read from $V_{\alpha'KKLT}$, but now expanding according to the LVS assumption that $W_0 \sim 1$ and that the minimum will be in the region of the moduli space where $a_s \tau_s \sim \log\mathcal{V}$.}
\begin{equation}
V_{LVS} = \frac{4}{3}\frac{e^{-2 a_s \tau_s}\sqrt{\tau_s} \ a_s^2 A^2 }{s\ \mathcal{V}}  - \frac{2 e^{- a_s \tau_s} \tau_s \ a_s A\ W_0 }{s\ \mathcal{V}^2} + \frac{3 \sqrt{s}\ \xi\ W_0^2}{8\ \mathcal{V}^3 }
\end{equation}
and where the contributions  $\Theta_{up}$ and $\Theta_{\alpha'/up} $ to the scalar masses are given by (\ref{mass1}) and (\ref{mass3}). The term  $\Theta_{\alpha'}$ is instead now given by 
\begin{equation}
\Theta_{\alpha'} = \frac{5 \sqrt{s}\ \xi\ W_0^2}{8 \mathcal{V}^3}(3 \alpha_1 -1)\:.
\end{equation}

In the LVS case, the minimum is non-supersymmetric already before adding the dS uplifting term. Minimising the LVS potential $V_{LVS}$, one obtains the following conditions that the K\"ahler moduli have to satisfy in the minimum:
\begin{equation}
e^{-a_s \tau_s} = \frac{3 \ \tau_s^{3/2} \ W_0}{a_s \tau_s \ A \ \mathcal{V}} \frac{a_s \tau_s -1 }{4a_s \tau_s -1}
\label{minlvs1}
\end{equation}
and 
\begin{equation}
\tau_s^{3/2} = \frac{s^{3/2} \xi}{2}\frac{1}{16 a_s \tau_s} \frac{(4a_s \tau_s -1)^2}{a_s \tau_s -1 } = \frac{s^{3/2} \xi}{2}\left( 1 - \frac{1}{16 a_s \tau_s} + \frac{9}{16}\frac{1}{a_s \tau_s -1}      \right)\:.
\label{minlvs2}
\end{equation}
This minimum is producing the cosmological constant term 
\begin{equation}
 V_0^{LVS} = -\frac{3 \sqrt{s}\ \xi \ W_0^2}{16 a_s \tau_s \ \mathcal{V}^3} = -3 \frac{s^{3/2} \xi}{4 \mathcal{V}}\frac{1}{a_s \tau_s} m_{3/2}^2\:.
 \label{CCseq}
\end{equation}
The non-zero value comes from the fact that the perturbative and non-perturbative corrections to the potential breaks its no-scale structure. Without these corrections, the tree-level potential would be zero at the minimum (but the K\"ahler moduli would be flat directions).
At the non-supersymmetric minimum, $F_{T_b} \neq 0$ generates a term in the supergravity scalar potential which goes like $F^{T_b} F_{T_b}$. Such a term at leading order cancels $3 m_{3/2}^2$ and is mainly responsible for the cancellation of the tree level cosmological constant in LVS. This does not happen in KKLT because the minimum is supersymmetric and therefore $ F_{T} = 0$. This difference is important because it will produce a difference in the parametric scaling  of the warp factor $\rho$ with volume $\mathcal{V}$ after imposing the dS/Minkowski condition.

We now  minimise the potential (\ref{Vtotlvs}) that includes also the $X$-contribution. The minimum condition  (\ref{minlvs1}) is not modified, while  (\ref{minlvs2}) is changed to
\begin{equation}
 \tau_s^{3/2} = \frac{1}{16 a_s \tau_s} \frac{(4a_s \tau_s -1)^2}{a_s \tau_s -1 } \left(\frac{s^{3/2} \xi}{2}  + \frac{8}{27 \beta_0} \frac{\rho^2\ \mathcal{V}^{5/3} }{W_0^2} \right)\:.
\label{minlvsnil}
\end{equation}
The dS/Minkowski condition
\begin{equation}
 V_0^{LVS} + V_{\alpha'up} = 0 
\end{equation}
restricts $\rho$ such that
\begin{equation}
 \rho^2 = \frac{27 \beta_0}{8} \frac{s^{3/2}\ \xi\ W_0^2}{\mathcal{V}^{5/3}(5a_s \tau_s -2)}\:.
 \label{CClvs}
\end{equation}
This produces a shift in the condition (\ref{minlvs2}):
\begin{equation}
 \tau_s^{3/2} = \frac{s^{3/2} \xi}{2}\left( 1 - \frac{3}{16} \frac{1}{5a_s \tau_s-1} + \frac{15}{16}\frac{1}{a_s \tau_s -1}      \right)\:.
\label{minlvs3}
\end{equation}
Moreover, we notice that if we plug the equation (\ref{minlvsnil}) into (\ref{minlvs1}), we obtain the relation
\begin{equation}
 e^{-a_s \tau_s} = \frac{3}{4 A}\frac{1}{a_s \tau_s}\left( \frac{s^{3/2} \xi}{2}\frac{W_0}{\mathcal{V}} + \frac{8}{27\beta_0}\frac{\rho^2 \mathcal{V}^{2/3}}{W_0} \right)\:.
 \label{lvscondall}
\end{equation}

Finally, if we introduce the dS LVS minimum in the scalar potential (\ref{Vtotlvs}) we can read the scalar mass. In particular we have
\begin{equation}
\Theta_{up} = \frac{9\sqrt{s}\ \xi}{16 \mathcal{V}^3}\frac{W_0^2}{5a_s \tau_s -2}\left( 1- \frac{s^{3/2}\xi}{\mathcal{V}}(1-\beta_1) \right)\left( 1-\frac{3\gamma_0}{\alpha_0 \beta_0}\right)\:,
\end{equation}
\begin{equation}
\Theta_{\alpha'/up} = \frac{9}{16}\frac{s^2 \xi^2 W_0^2}{\mathcal{V}^4}\frac{3\gamma_0}{\alpha_0 \beta_0}\frac{1}{5a_s \tau_s-2}\left( \gamma_1 -\alpha_1 -\beta_1  \right)
\end{equation}
and 
\begin{equation}
\Theta_{\alpha'} = \frac{5 \sqrt{s}\ \xi\ W_0^2}{8 \mathcal{V}^3}(3 \alpha_1 -1)\:.
\end{equation}
Therefore, in this minimum the dominating contributions are $\Theta_{\alpha'}$ and the first term of $\Theta_{up}$. However if the K\"ahler potential has the log structure \eqref{paramK\"ahleralpha}, with the parameters given by (\ref{log2}), there is a cancellation in $\Theta_{up}$ which makes it vanish. Therefore the scalar masses would be given in this case by $\Theta_{\alpha'}$.

\subsection{Soft terms and F-terms in LVS with nilpotent goldstino}
\subsubsection{F-terms}
In the dS LVS minimum, susy breaking is dominated by the F-term of the modulus $T_b$ determining the volume (we will call it $F^{\vo}$ instead of $F^{T_b}$ to make this clear):
\begin{equation}\label{FvolLVS}
 F^{\mathcal{V}} = -2\, \mathcal{V}^{2/3}\, m_{3/2} - \frac{s^{3/2}\, \xi}{24\,  \mathcal{V}^{1/3}}\frac{80 a_s^2 \tau_s^2 -67 a_s \tau_s +32}{(5a_s\tau_s -2)(a_s \tau_s -1)}\, m_{3/2}\:,
\end{equation}
where the first  term cancels the $-3m_{3/2}^2$ term in the scalar potential at leading order, due to the underlying no-scale structure.
Expanding \eqref{FvolLVS} in powers of $\frac{1}{a_s \tau_s} \sim \frac{1}{\log\mathcal{V}}$, we obtain at leading order
\begin{equation}
 F^{\mathcal{V}} = -2\, \mathcal{V}^{2/3}\, m_{3/2} - \frac{2}{3}\frac{s^{3/2}\, \xi}{ \mathcal{V}^{1/3}}\,  m_{3/2} \:.
\end{equation}
The F-term of the nilpotent goldstino,
\begin{equation}
F^X = -\frac{2}{3}\sqrt{\frac{3}{2\beta_0}}\sqrt{\frac{s^{3/2}\, \xi}{5a_s\tau_s-2}}\frac{1}{\mathcal{V}^{1/6}}\,  m_{3/2}\:,
\end{equation}
is subleading with respect to the $\mathcal{V}$ modulus F-term \eqref{FvolLVS}. Even the small modulus $T_s$ has a bigger susy breaking contribution through the F-term
\begin{equation}
F^{T_s}=- \frac{6\tau_s}{4a_s\tau_s-1}\,  m_{3/2}\:.
\end{equation}

The dilaton contribution to susy breaking is the smallest one. As in KKLT, its F-term can receive contributions from both $D_TW$ and $D_SW$. The last one can be parametrised as in \eqref{dilatonmin}, for the same reason explained in section \ref{FtermsKKLT}. For LVS, this contribution is of the same order of the one coming from $D_TW$. If we expand in $\frac{1}{a_s \tau_s} \sim \frac{1}{\log\mathcal{V}}$, we obtain 
\begin{equation}
F^S = \frac{3}{2}\frac{s^{5/2}\, \xi}{\mathcal{V}}\, \left( 3-2\omega_s\right) \, m_{3/2}
\end{equation}
where again $\omega_s\lesssim 1$.

\subsubsection{Soft terms}
In this section we are going to discuss the soft terms for a visible sector living on D3-branes, with moduli stabilised at the LVS dS minimum. 

The scalar masses already discussed in the last section can be written at leading order as:
\begin{equation}
 m_0^2 = \frac{5}{4} \frac{s^{3/2} \, \xi}{\mathcal{V}} \, (3 \alpha_1 -1)\, m_{3/2}^2 + \frac{9}{8}\frac{s^{3/2} \, \xi}{\mathcal{V}} \, \frac{1}{5a_s \tau_s} \, \left(1-\frac{3 \gamma_0}{\alpha_0 \beta_0} \right) m_{3/2}^2 \:.
\end{equation}
 Assuming the {\it log hypothesis} for the K\"ahler potential, i.e. imposing the relations (\ref{log2}) in the effective K\"ahler potential (\ref{K\"ahlerlvs}), the scalar masses are given by
\begin{equation}\label{LVSsoftScM}
m_0^2 = \frac{5}{8}\frac{s^{3/2}\xi}{\mathcal{V}}m_{3/2}^2
\end{equation}
which are completely dominated by the $\alpha'$ contribution. Regarding the gaugino masses, as discussed in \cite{bckmq, ackmmq}\,  the uplift term plays only an indirect role. 
The gauge kinetic function is $f\simeq S$ in the case of branes at singularities.
This implies that the gauginos are completely dominated by $F^S$. Hence the expression  for the gaugino masses at leading order in $\frac{1}{a_s\tau_s}\sim\frac{1}{\log\mathcal{V}}$ is
\begin{equation}\label{M12LVSb}
M_{1/2} =\pm \frac{3}{4}\frac{s^{3/2}\xi}{\mathcal{V}}\left( 3-2\omega_s \right)\,m_{3/2} \:,
\end{equation}
where the relative sign $\pm$ refers to the choice of $W_0\gtrless 0$. 
Finally the trilinears at leading order in $\frac{1}{a_s\tau_s}\sim\frac{1}{\log\mathcal{V}}$ can be written in terms of the gaugino mass as
\begin{equation}
A_{ijk} = - (1- s\partial_s \log Y_{ijk}^{(0)}) M_{1/2} \:.
\end{equation}

Notice that the general structure of soft terms is similar to the one found in \cite{ackmmq} for the local case in which the uplift term is given by hidden sector matter fields. The volume suppresion of the soft terms with respect to the gravitino mass is a sign of sequestering. 
The potential sources of de-sequestering are discussed in Appendix B of \cite{ackmmq}, where it is shown that their effects are irrelevant for D3-branes at singularities with the potential exception of the effect of field re-definitions induced by quantum corrections to gauge couplings, as computed in \cite{Conlon:2009kt}\footnote{Notice that this field re-definition refers to the visible sector blowing-up mode. In the hidden sector 
the anti-D3-brane is on top of an orientifold O3-plane (with no orbifold singularity in the double cover Calabi-Yau);
its massless spectrum corresponds only to the goldstino with no gauge fields and therefore the analysis of \cite{Conlon:2009kt} does not apply.}. If this redefinition is naively substituted in the K\"ahler potential it may give rise to de-sequestering, as discussed in \cite{Conlon:2010ji}. The effect of these field re-definitions in the K\"ahler potential has not been computed explicitly so this remains as an open question. The same considerations apply to our models.

\section{Some cosmological and phenomenological observations}\label{Sec:CosmoPhenoObserv}

In this section we consider the mass of the lightest modulus and its relation to the relevant scales in both scenarios. 
We will discuss possible cosmological consequences  and how they constrain the scenarios discussed in this paper. For instance, as we will see, 
the SUSY breaking scale develops some bounds. Finally we will speculate on slow-roll inflation. 
The results presented in this section hold strictly for an MSSM visible spectrum. However several features will still be valid for close modifications of the MSSM.

 \subsection{LVS with D3-branes}
 The lightest modulus in LVS is the volume modulus. Its mass (that as usual is computed from the matrix of the second derivatives of the potential) is, in terms of the gravitino mass,
 \begin{equation}
  m_{\mathcal{V}}^2 = \frac{45}{8}\frac{s^{3/2}\xi}{\mathcal{V}}\frac{20 a_s^3\tau_s^3 -21 a_s^2\tau_s^2 +9a_s \tau_s -2}{\left(8 a_s^3\tau_s^3 -6 a_s^2\tau_s^2 +3a_s \tau_s +1 \right)(5_s\tau_s -2)}\, m_{3/2}^2\:.
 \end{equation}
At leading order in  $\frac{1}{a_s \tau_s} \sim \frac{1}{\log{\mathcal{V}}} \ll 1$ expansion, one obtains
\begin{equation}
  m_{\mathcal{V}}^2 = \frac{45}{16}\frac{s^{3/2}\xi}{a_s \tau_s \, \mathcal{V}}\, m_{3/2}^2\:.
  \label{lightmoduluslvs}
\end{equation}
Comparing this with \eqref{LVSsoftScM}, one can conclude that there is the following hierarchy between the relevant scales:
\begin{equation}
  m_{3/2} > m_0 > m_{\mathcal{V}}\:.
\end{equation}
Since the lightest modulus redshifts like the matter does, 
it quickly dominates the thermodynamic history of the universe after the end of  inflation \cite{CMP,NTDM}. Through its decay, it reheats the universe, but
being its mass smaller than $m_{3/2}$, it
is not able to produce gravitini through direct decay. Hence, in this scenario there is no gravitino problem.

The volume modulus can decay into SM (MSSM) particles. Since the moduli couple to matter gravitationally, 
the lightest modulus decays very late and that could in principle spoil nucleosynthesis. 
One way of quantifying it is through the decay (reheating) temperature of this modulus, which is given by $T_{RH} \propto \sqrt{\Gamma \, M_p}$, where $\Gamma$ is the decay rate and $M_P = 2.4 \ 10^{18}$ GeV. Since in this case $\Gamma\sim \frac{m_\vo^3}{M_p^2}$, the reheating temperature is
\begin{equation}
 T_{RH} \simeq \sqrt{\frac{m_{\mathcal{V}}^3}{M_P}} \:.
 \label{temp}
\end{equation}
 In order to avoid problems with nucleosynthesis one should have $T_{RH} \gtrsim 4$ MeV \cite{Hannestad:2004px}. This would impose a bound on the lightest modulus mass. This is the so called Cosmological Moduli Problem (CMP), that is known to affect the LVS scenario if the soft masses are at the TeV scale.
 At the same time it would impose a bound on the gravitino through the relation (\ref{lightmoduluslvs}). Finally given the relation \eqref{LVSsoftScM} between the gravitino and the scalar mass, this bound on the volume modulus implies a bound on the scalar masses and hence on the susy breaking scale. 

We can also make use of \eqref{lightmoduluslvs} and \eqref{LVSsoftScM} to write the volume modulus in terms of the scalar masses. 
When the visible sector lives on  D3-branes, 
we have 
\begin{equation}\label{m0mmodhier}
 m_{\mathcal{V}}^2 = \frac{9}{2} \frac{1}{a_s \tau_s}\, m_0^2 \:.
\end{equation}
Using the bound on the reheating temperature one can see that for typical (GUT) values $s\sim 10$ and $\mathcal{V}\sim 10^6 - 10^7$  the scalar masses are forced to be bound as
\begin{equation}
 m_0 \gtrsim 65\ \mathrm{TeV}
\end{equation}
and therefore $M_{\rm SUSY}\gtrsim 65$ TeV\footnote{$M_{\rm SUSY}$ refers to the scale of the scalars, typically the scale of the stop.}. Moreover, in this scenario one has a hierarchy between the scalar and the gaugino masses (a split-like scenario \cite{Giudice:2004tc,ArkaniHamed:2004yi,Wells:2004di}). Actually, for the same values of the dilaton $s$  and the volume $\mathcal{V}$ that we used before, we have
\begin{equation}\label{m0m12hier}
 m_0 \simeq (10^2 - 10^3) M_{1/2}\:.
\end{equation}
One could then in principle work out a scenario in which the electroweakinos are at the TeV scale, that would make them potentially interesting as dark matter candidates. 
The equation \eqref{M12LVSb} implies universality of the gaugino masses at high energies. 
For this reason, following the RG-flow, the lightest gaugino can only be the bino. Depending on the electroweak symmetry breaking conditions on the MSSM, the lightest  electroweakino will then be either the bino or the higgsino. 

The thermal averaged cross section $\langle \sigma v \rangle$ for the bino annihilation is very small, hence it is very difficult not to overproduce dark matter bino like unless there exists a co-annihilation with other sparticles. Given the hierarchy \eqref{m0m12hier} between scalars and gauginos, co-annihilation with sleptons or the A-funnel would not be realisable. 

Notice that in more complicated supersymmetric models there could be more neutralino components (e.g. the singlino one in the NMSSM). However for the analysis considered below this fact does not produce any relevant difference. Going beyond this will need a more involved analysis that goes beyond the scope of this paper.

Hence the only option is a scenario where the dark matter is a neutralino that is higgsino like or a bino like one which co-annihilates with a NLSP higgsino. In this case, such a scenario is possible for scalars in the range
\begin{equation}
 10^5\ \mathrm{TeV} \gtrsim m_0 \gtrsim 65\ \mathrm{TeV}\:,
\end{equation} 
where the upper bound is a consequence of the hierarchy \eqref{m0m12hier}: 
if scalars were heavier than this bound, the gauginos would be heavy enough to induce a  one loop contribution to the higgsino mass bigger than 1 TeV. Such a heavier higgsinos would overproduce dark matter. On the other hand if the scalars were heavier than $10^3$ TeV, following the same hierarchy, the binos would be very heavy and the bino-higgsino scenario would not be possible.

Notice that for  $m_0\lesssim 10^4$ TeV, \eqref{m0mmodhier} implies that the mass of the modulus is below $5000$ TeV and therefore the reheating temperature \eqref{temp} would be below $7.5$~GeV. Then, any neutralino with a mass heavier than 150~GeV would have freeze-out temperature $T_{fo}\simeq \frac{m_\chi}{20}$ above the reheating temperature $T_{RH}$. Hence, its relic abundance would be produced non-thermally. Interestingly, non-thermally produced higgsinos could saturate the total relic density even if their mass is below 1~TeV (for similar discussion see \cite{acdkmq}).

The typical pattern of masses in this scenario is the following: there are near degenerate neutralinos $\chi_1^0$,  $\chi_2^0$ and a chargino $\chi_1^\pm$ with masses around 1~TeV, whereas the rest of the electroweakinos are heavier and the scalars are in the multi TeV range (a little split in this case). In the case of pure higgsino the collider phenomenology is dominated by hard jet production with large missing energy which is known as monojet search. There has been a lot of work in this direction. In particular the authors in \citep{Low:2014cba}  claim that the exclusion limits for higgsino masses at LHC 14~TeV are around 185~GeV and at a 100~TeV machine would reach the 870~GeV (both for luminosity $\mathcal{L}\sim 3000\, fb^{-1}$). 

In the case of bino like neutralino, if the split between the bino and the higgsino  were around 20-50~GeV, then higgsinos could decay into binos via off-shell gauge bosons which could produce a signal with low $p_T$ leptons.\footnote{Notice that this signal is different from the  standard multilepton one which is produced through squark decay. In this scenario, given that scalars are very heavy, LHC will not be able to produce them and hence the standard multilepton signals do not apply.} The exclusion limit for bino masses at LHC 14 (at $\mathcal{L}\sim 3000\, fb^{-1}$) is around 300~GeV, and in a future 100~TeV collider would be around 1~TeV (for the same luminosity)~\citep{Low:2014cba}.

Dark matter direct detection experiments will shed light on bounds on electroweakino masses. In our case, these experiments will  have a definite impact only if the gaugini are light enough, i.e. $M_{1/2} \lesssim 20$ TeV. Due to the hierarchy \eqref{m0m12hier}, that would be possible only if the scalars were lighter than $10^4$~TeV.
Otherwise, for example for the pure higgsino, if the binos were heavier than 10 TeV,  all the parameter space would escape the bounds coming from direct detection experiments  \citep{diCortona:2014yua, Badziak:2015qca}. The reason is that $\mu < M_1$ and then the spin independent cross section goes like $\sigma_{SI}\sim 1/M_1^2$, i.e.  a bigger $M_1$ corresponds to a smaller  cross section.

The case of mixed bino/higgsino needs $\sim 1$ TeV gauginos, but that would imply scalar masses $\sim 10^3$ TeV. Such scalars allow a 125 GeV higgs in the region where $tan\beta\sim 2$ \cite{Vega:2015fna}. For this value of $\tan\beta$ and $\mu<0$, the spin independent cross section is below the strongest limits on direct detection which so far are given by  LUX  \cite{Akerib:2013tjd}. The new limits  by XENON 1T are expected for next year and will be very sensitive for this scenario.

Concerning dark matter indirect detection,  the strongest bounds for higgsino and bino/higgsino  dark matter come from $\gamma$-rays produced by neutralino annihilation. In particular the most stringent ones  come from Fermi-LAT's data on dwarf Spheroidal Galaxies \citep{Ackermann:2015zua}. However these limits are not decisive for the two scenarios discussed in this section. Future experiments like CTA \cite{Carr:2015hta} will have a bigger impact on higgsino and bino/higgsino mass limits.

\subsection{KKLT with D3-branes}   \label{Sec:KKLTpheno}
The lightest modulus in KKLT is the K\"ahler modulus $T$. Its mass is given by
\begin{equation}
 m_T = 2 a\, \mathcal{V}^{2/3}\, m_{3/2}
\end{equation}
where $\mathcal{V}=\tau^{3/2}$. This time the relation between the relevant scales is
\begin{equation}\label{massRelKKLT}
  m_{T} > m_{3/2} \gtrsim m_{0}\:.
\end{equation}
The last relation depends whether the {\it log hypothesis} is realised or not: in the first case 
the soft scalar masses are suppressed with respect to the gravitino mass,
while in the second case the two masses are of the same order.

Given the structure of \eqref{massRelKKLT}, there will be no cosmological moduli problem because the lightest modulus is heavier than the visible sector scalars. However, the fact that the modulus is heavier than the gravitino can lead to a moduli-induced gravitino problem. That happens because now the channel $T\rightarrow 2\psi_{3/2}$ is no longer closed. In fact its decay rate is generically large \cite{Endo:2006zj,NakamuraYama}. 

The gravitinos are produced by direct decay of the modulus after inflation.  However this has non trivial cosmological consequences: the requirement that the gravitino decay products should not spoil the nucleosynthesis constrains strongly the gravitino abundance putting a bound on its mass, i.e. $m_{3/2}\gtrsim 10^5$ GeV. However, the gravitino can decay to R-parity odd particles, like stable electroweakinos. That could overproduce relic density of gauginos or higgsinos. This generates a more severe bound:  if the LSP is wino like one has $m_{3/2}\gtrsim 10^6$ GeV, while for higgsinos and binos it is stronger, i.e. $m_{3/2}\gtrsim 10^7$ GeV. 

The bound on the gravitino mass translates into a bound on the scalar masses. When the K\"ahler potential takes the generic form \eqref{KparamEFT}, $m_0\sim m_{3/2}$ and one can read the bound from the previous paragraph. If the {\it log hypothesis} is satisfied, one obtains for the scalar masses:
\begin{equation}
 m_0 \gtrsim \sqrt{\frac{s^{3/2}\xi}{4\mathcal{V}}}\,(10^3 - 10^4) \mathrm{TeV}\:,
\end{equation}
which, for values of dilaton $s\sim 10$ and volume $\mathcal{V}\sim10^3$, becomes 
$
 m_0 \gtrsim 9- 900\, \mathrm{TeV}
$. 
9 TeV correspond to winos and 900 TeV to the bino case (higgsinos would be somewhere in between). 

On the other hand, in KKLT the anomaly mediation contribution to gaugino masses (see Appendix \ref{Sec:AnomalyMed}) is of the same order as the moduli mediated one. Therefore, using the expression for soft-terms discussed in Section \ref{Sec:KKLT} and in Appendix \ref{Sec:AnomalyMed} we see that
\begin{equation}
 M_a^{KKLT} =  \left( \frac{3}{2}\frac{1}{a\mathcal{V}^{2/3}} -\frac{g_a^2}{16 \pi^2} b_a \right)  \,  m_{3/2}\:,
 \label{mirage}
\end{equation}
where $g_a^2 \simeq 4\pi /s$. Following the notation of the equation (3.1) in Choi, Jeong and Okumura \cite{cfno,cfno2,cfno3}, one could rewrite it as 
\begin{equation}
 M_a^{KKLT} = \frac{3}{2}\frac{1}{a\mathcal{V}^{2/3}}  \left( 1 -\frac{a \mathcal{V}^{2/3} g_{GUT}^2 }{16 \pi^2}  b_a \, \hat\alpha \right)  \,  m_{3/2}\:,
\end{equation}
where according to them $a \mathcal{V}^{2/3}=\log(M_P/m_{3/2})$. By doing this, one can see that  
\begin{equation}
\hat\alpha \equiv \frac{2}{3}\frac{1}{g_{GUT}^2}\frac{4\pi}{s}\:.
\end{equation}
The mirage scale (which is the scale at which the gauginos unify) will be given by
\begin{equation}
M_{mir}=M_{UV}\, e^{-a\mathcal{V}^{2/3}\frac{\hat\alpha}{2}}\:.
\label{mirage2}
\end{equation}
For $s=10$ one has $\hat\alpha \simeq 1.7$ and more interestingly, for $s=8.5$ ($g_s\simeq0.12$) one has $\hat\alpha =2$. Therefore,  from \eqref{mirage2} for a volume of $\mathcal{V}\sim 10^3$ and given that $a=2\pi/N$ the scale can be written as  
\begin{equation}
M_{mir}=M_{UV}\, e^{-\frac{200 \pi}{N}} \:.
\label{mirage3}
\end{equation}
Hence, the behaviour is now dominated by the number $N$. For $N$ small the scenario is anomaly mediation dominated. When $N$ is very large, the scenario is modulus dominated. There is a particularly interesting case: for $N\simeq 21$  the mirage scale is at the TeV scale (when $M_{UV}\sim 10^{16}$~GeV). This scenario is reproducing the one studied in \cite{cfno,cfno2,cfno3} and \cite{Choi:2006im} where the pattern of masses at the TeV corresponds to a compressed spectrum scenario. These scenarios have as dark matter  candidates higgsino like neutralinos or a mixture of near degenerate bino/wino/higgsino.

As it will be discussed in Appendix \ref{Sec:AnomalyMed}, in KKLT the anomaly mediation contributions for sleptons are negative. 
This is not a problem when the {\it log hypothesis} is not fulfilled, as they are suppressed by a loop factor with respect to the gravitino mass. On the other hand, if this hypothesis is realised, 
one has a bound to avoid tachyonic scalar squared masses: the highest anomaly mediation contribution is in $m_{\tilde{e}_R}$; summing this with the contribution discussed in Section~\ref{Sec:KKLT}, one obtains
\begin{equation}
m_i^{2}\lvert_{KKLT} =  \left(\frac{s^{3/2}\xi}{4\mathcal{V} } -\frac{8 g_a^4}{(16 \pi^2)^2}  \right)  \,  m_{3/2}^2\:,
\end{equation}
giving the bound on the volume
\begin{equation}
\mathcal{V} < \frac{s^{9/2}\xi \pi^2}{2}\simeq 10^5\:,
\end{equation}
where we used again $g_a^2 \simeq 4\pi /s$. 

Notice that for $N<20$ (and also in the case of $\omega_s <<1$) the anomaly mediation contribution dominates and then the phenomenology changes. In this case an open window to wino like dark matter is opened. Moreover, depending on the higgsino mass, dark matter matter could also be  higgsino like. Even scenarios with co-annihilating  wino-higgsino or  wino-bino would be allowed. For the case of wino like dark matter,  disappearing tracks search is competitive with the monojet one \citep{Low:2014cba}. The exclusion limits on wino masses at LHC 14 TeV are around 280 GeV and at a 100 TeV machine would reach the 2.1 TeV (both for luminosity $\mathcal{L}\sim 3000\, fb^{-1}$). The limits on the co-annihilating bino-wino or wino-higgsino are similar to those discussed in the previous sub-section for the bino-higgsino case. 

The impact of dark matter direct detection on the anomaly mediated scenario has been studied in \citep{diCortona:2014yua, Badziak:2015qca}.  It depends very much on the higgsino mass but the combination of dark matter direct detection and collider searches seems to be a powerful tool. Unfortunately, to really constrain the parameter space a 100 TeV machine is needed. Finally, in the indirect detection searches of dark matter for wino like WIMPs there is a discussion on possible exclusion limits on thermal winos (see for example \cite{Hryczuk:2014hpa}). The bounds come from signals of monochromatic photons from our Galactic Core coming from possible neutralino annihilation, in particular the monochromatic H.E.S.S. line~\cite{Abramowski:2013ax}.

   \subsection{Slow-roll inflation and nilpotent goldstino}

Models of inflation in string theory abound \cite{bcq,bm}. 
A usual criticism to these models is the fact that they assume the presence of the uplift term without specifying its source. Moreover, if supersymmetry is broken explicitly by the uplift term one could doubt that  the corresponding field theory is under control. With the formalism used in this paper, one realises the uplift term by introducing the nilpotent superfield. This automatically provides
a concrete supersymmetric description of the inflationary models. 
This might be applied to models present in the literature like, for instance, the K\"ahler \cite{kahlerinflation} and fibre moduli inflation \cite{fibre}: here the inflationary region behaves as $V\sim A-B e^{-k\phi}$ with $A$ and $B$ independent of the inflaton $\phi$ and $A$ determined by the uplift term. 
For  recent proposals of inflation in supergravity models along the lines described here see for instance \cite{adfs, Ferrara:2014kva, Kallosh:2014via, Galante:2014ifa, Dall'Agata:2014oka, Dudas:2015eha, Carrasco:2015uma, dudaswieck, scalisi}.

Furthermore, notice that finding the general structure of soft scalar masses for D3-branes is essentially the same calculation needed for the well studied brane/anti-brane inflation scenario. In \cite{kklmmt,postkklmmt} several contributions to scalar masses were studied in order to compensate for the  $-2V_0/3$ contribution that gives rise to the $\eta$ problem. 
The $-2V_0/3$ contribution
appears also in our formulae \eqref{ScPotKKLTUp} and \eqref{Vtotlvs}, where, during infation,
 $V_0$ is not equal to zero, but it is positive. On the other hand, in our case the other contributions to the scalar masses are not proportional to $V_0$.
It is then conceivable to tune the parameters such that the $2V_0/3$ contribution is approximately canceled in the quadratic term (in the inflaton) of the potential, giving rise to slow roll inflation. 
This can be combined with the other supersymmetric contributions described in \cite{kklmmt,postkklmmt} in order to estimate the required $\mathcal{O}(10^{-2})$ fine tuning.

\section{Summary and conclusions}\label{Concl}
In this article we have considered a particular 4D supergravity effective field theory with a nilpotent superfield, in which the supersymmetry is realised non-linearly. Following \cite{kw,kw2}, we have argued that this should be the low energy effective theory describing the moduli and matter physics of CY flux compactifications of type IIB string theory with an anti-D3-brane at the tip of a warped throat. 

We summarise our findings as follows.
\begin{enumerate}
\item{}
The coupling of the nilpotent superfield $X$ to moduli and chiral matter provides the uplifting term proposed in KKLT \cite{kklt,kw,kw2}. If $X$ couples to the moduli  in the K\"ahler potential in the same way as the D3-brane matter superfield $\phi$, then the generated de Sitter uplift term has the same CY volume dependence as the one coming from an anti-brane at the tip of a warped throat \cite{kklmmt}. 
Adding a coupling between $X$ and $\phi$ in the superpotential, we could  reproduce  also the brane/anti-brane Coulombian potential described for instance in \cite{kklmmt}.

\item{} 
The anti-D3-brane is taken to be bound on top of an O3-plane, that is placed at the tip of the throat. This does not allow the anti-D3-brane to move. 
On the other hand we want to realise the visible sector on a set of D3-branes placed at a singularity. The fluxed induced soft masses typically stabilise the position of these D3-branes. On the other hand, the anti-D3-brane attracts the D3-brane to the throat. This may in principle destabilise the system, moving the D3-brane outside the singularity (in this case the SM gauge group would be destroyed). We checked how big the soft masses must be such that this does not happen. We
found that in the studied cases the system is stable, with the exception of the so-called ultralocal sequestered scenario in LVS. 

\item{}
We analysed the structure of the de Sitter vacuum in KKLT and how the $F$-term of the nilpotent field $X$ induces soft supersymmetry breaking terms for D3-branes at singularities.
We found that if the K\"ahler potential can be brought in the logarithmic no-scale form \eqref{Klog}, the soft scalar masses vanish at leading order. Only when $\alpha'$ effects are included these soft terms are non-vanishing, but suppressed with respect to the gravitino mass (see Table \ref{Tab:D3}).
As discussed in Section~\ref{Sec:KKLTpheno}, in this case the anomaly mediation contribution can compete (possibly inducing tachyonic masses).
On the other hand, if the {\it log hypothesis} is not realised, 
sfermion masses are of order $m_0\sim m_{3/2}$, while the other soft terms are of order $m_{3/2}/\log\left(M_p/m_{3/2}\right)\sim \mathcal{O}\left(10^{-2}m_{3/2}\right)$. As regard the gaugino masses, typically the anomaly mediation contribution dominates.
As we have explained in the introduction, this scenario has some analogies with the one found by other means in \cite{cfno,cfno2,cfno3,Lebedev:2006qq} and extends the results of  \cite{kqu} to include $\alpha'$ corrections.

\item{}
We studied for the first time the explicit structure of soft terms induced by an anti-D3-brane in the Large Volume Scenario (LVS). 
We described the anti-D3-brane uplift by introducing the nilpotent field like in KKLT. We computed the structure of soft terms in this case as well. We found a concrete realisation of split supersymmetry in which TeV gaugino masses $M_{1/2}$ are lighter than the scalar ones $m_0$ by a factor $\vo^{-1/2}$. Moreover, the scalars are lighter than the gravitino by the same factor, with $m_0^2\sim M_{1/2}m_{3/2}$. In order to have a TeV gaugino the volume must be of order $\vo\sim 10^{6}-10^{7}$ \cite{bckmq,ackmmq}.\footnote{For LVS, the {\it log hypothesis} does not play any role, as it induces cancellations in the subleading contributions.} Notice that the used formalism allows to treat both sources of supersymmetry breaking at the same level. 
The dominant component comes from the overall volume modulus but all sources of supersymmetry breaking play a role due to standard no-scale cancellations. This scenario gives the same physics as those  obtained in \cite{bckmq,ackmmq} in which other uplifting mechanisms were used. 

\item{}  
We have commented some possible phenomenological consequences of the KKLT and LVS scenarios with nilpotent goldstinos. In both cases the scalars are  heavier than gauginos such that the only possible accesible sparticles at TeV scales are some neutralinos and some charginos. It seems that LHC exclusion limits for electroweakinos are not decisive at all~\cite{Low:2014cba}. Hence a 100 TeV machine would be desirable to explore the most interesting corners of their parameter space. We have also made some comments on the possible impact coming from dark matter direct and indirect detection. The LVS scenario behaves as a mini-split susy model with higgsino-like or bino higgsino as dark matter candidates. In the KKLT scenario, the scalars are a bit heavier than gauginos and the dark matter candidates depend on how much anomaly mediation dominates. On the one hand, it could have a compressed spectrum with dark matter being higgino like or a mixture higgsino-bino.
Alternatively it could be anomaly dominated and then, also wino like dark matter would be possible.
\end{enumerate}

We summarise the structure of soft terms for matter on D3-branes for both KKLT and LVS in Table \ref{Tab:D3}, under the assumption that the K\"ahler potential takes the logarithmic form \eqref{Klog}.\footnote{Notice that the soft terms are non-vanishing only when non-perturbative effects, $\alpha'$ corrections and the presence of the nilpotent superfield are considered. This is consistent with the existence of a vanishing supertrace formula recently found in \cite{bena} since in that reference those effects were not included.} 
In summary, including also the study of the visible sector living on D7-branes presented in Appendix \ref{App:D7-brane} and summarised in table \ref{Tab:D7}, there are four distinct scenarios, depending whether the visible sector lives on D3 or D7-branes and on  the moduli stabilisation mechanism (KKLT or LVS).
These may be subject to strong constraints in the not too far future by LHC and its potential extensions and different dark matter searches.

  \begin{table}
\begin{center}
{\tabulinesep=1.4mm
   \begin{tabu}{|c|c|c|}
\hline
 & {\bf KKLT} &  {\bf LVS} \\
 \hline\hline
Soft term &  D3 &  D3   \\ \hline \hline
$M_{1/2}$ & $\pm\left(\frac{3}{2a\vo^{2/3}}   \right)\, m_{3/2}$ & $\pm\left(\frac{3s^{3/2}\xi}{4\vo}\right)\, m_{3/2}$   \\ \hline
$m_0^2$ & $\left(\frac{s^{3/2}\xi}{4\vo}\right) m_{3/2}^2$ &  $\left(\frac{5s^{3/2}\xi}{8\vo}\right)\, m_{3/2}^2$  \\ \hline
$A_{ijk}$ & $-(1-s\partial_s\log Y_{ijk})\, M_{1/2} $& $-(1-s\partial_s\log Y_{ijk})\, M_{1/2} $\\ \hline
\hline

\end{tabu}}
\small\caption{Summary of different soft terms for the visible sector on D3  branes for both KKLT and LVS scenarios (when the {\it log hypothesis} is fulfilled). In both cases there is a hierarchy of masses with the ratio $\epsilon=M_{1/2}/m_0\ll 1$. For typical numbers we have $\epsilon\sim 1/50$ for KKLT and $\epsilon\sim 10^{-2}-10^{-3}$ for LVS, illustrating a  version of mini-split supersymmetry.}\label{Tab:D3}

\end{center}
\end{table}

One could generalise the use of the supersymmetric effective field theory with a nilpotent superfield in different setups.
For example, adding the effects of several anti-D3-branes in terms of several nilpotent superfields seems straightforward and may lead to richer scenarios. Moreover, here we have always assumed that the anti-D3-brane is on top of an O3-plane; if this is not the case, there will be other degrees of freedom that could be captured by including different constrained superfields. We hope the results of this article could be useful for further developments.

\section*{Acknowledgements}
We thank useful conversations with Kiwoon Choi, Michele Cicoli, Shanta de Alwis, Rajesh Gupta, Renata Kallosh, Sven Krippendorf, Anshuman Maharana, Luca Martucci, Joe Polchinski, Marco Serone, Angel Uranga.

  \appendix
  
  \section{Anomaly mediated contributions}\label{Sec:AnomalyMed}

Anomaly mediation  \cite{RandallSundrum,GiudiceLutyEtAl} generates a one-loop gaugino mass and two-loop scalar masses and is always present if there exists a hidden sector in the theory. The expressions for anomaly mediated contributions to scalars and gaugino masses are given by :
\begin{equation}
 M_{anom} = \frac{\beta_{g_a}}{g_a} m_{3/2} \qquad\mbox{and}\qquad
 m_i^{2}\lvert_{anom} = \frac{1}{2}\frac{d \gamma^i}{dt} m_{3/2}^2\:,
\label{anomscal}
 \end{equation}
where $\gamma^i$ is the anomalous dimension and $\beta_{g_a}$ is the beta function for the gauge couplings $g_a$. One can make more explicit the expression for the scalar masses in (\ref{anomscal})
\begin{equation}
  m_i^{2}\lvert_{anom} = \frac{m_{3/2}^2}{2} \left(\beta_{g_a} \frac{\partial}{\partial g_a} + \beta_{y^{kmn}} \frac{\partial}{\partial y^{kmn}} + \beta_{y^*_{kmn}} \frac{\partial}{\partial y^*_{kmn}} \right)\gamma^i \:,
\end{equation}
where $\beta_{y_{kmn}}$ is the beta function for the Yukawas. In particular, the expression for the anomalous dimension is
\begin{equation}
 \gamma^i = \frac{1}{16 \pi^2}\left( \frac{1}{2}\sum_{m,n}|y_{imn}|^2 -2 \sum_{a} g_a^2 C_a(i)\right) 
\label{anomdim} 
\end{equation}
where $ C_a(i) $ are the quadratic Casimir invariants of the group in the fundamental representation. The beta function for the gauge couplings in the MSSM are given by
\begin{equation}
 \beta_{g_a} = -\frac{g_a^3}{16 \pi^2}  (3T_G -T_R)
\label{betagaug}
 \end{equation}
where $T_G$ is the Casimir invariant in the adjoint representation and $T_R$ is the Dynkin index of the group. In the MSSM:

\begin{equation}
 3T_G - T_R =   \left\lbrace
  \begin{array}{c c l}
     -\frac{33}{5} &\text{ for }& U(1)_Y  \\
     -1 &\text{ for } & SU(2)_L \\
     +3 &\text{ for }& SU(3)_c \\
  \end{array}\:.
  \right.
  \label{anomparam}
\end{equation}
Finally the beta function for Yukawas can be written generically as
\begin{equation}
 \beta_{y_{ijk}} = \gamma^i_n y^{njk} + \gamma^j_n y^{ink} + \gamma^k_n y^{ijn}\:.
 \label{betayuk}
\end{equation}

From (\ref{anomscal}) and (\ref{betagaug}) one can read the anomaly mediated contribution to gaugino masses:
\begin{equation}
 M_a^{anom} = -\frac{g_a^2}{16 \pi^2} (3T_G -T_R) m_{3/2}
\label{anomgaug2}
 \end{equation}
and from (\ref{anomscal}) and (\ref{anomdim}-\ref{betayuk}) one could also see that the dominating contribution to the scalars is governed by the contribution of the gauge couplings 
\begin{equation}
   m_i^{2}\lvert_{anom} = \frac{2}{(16 \pi^2)^2} \sum_a g_a^4 C_a(i) (3T_G -T_R) m_{3/2}^2\:.
\label{anomscal2}
\end{equation}
From this expression we see that the sleptons in pure anomaly mediated susy breaking are tachyonic.

\

A  way of understanding anomaly mediation was proposed in ~\cite{Bagger:1999rd,thaler}\footnote{See \cite{shanta} for a different point of view.} as a susy preserving effect in AdS$_4$. In that case the authors propose that the AdS susy structure is the  underlying symmetry structure for SUGRA theories. In order to preserve such underlying AdS susy structure, it is needed that on top of the loop anomaly mediated terms described above, one has to take into account the one-loop goldstino couplings. That generates  general expressions for the soft terms in anomaly mediation for flat space of the form:
\begin{equation}
M_a^{anom} =  \frac{\beta_{g_a}}{g_a} \left(m_{3/2} - \frac{1}{3}K_l  F^l\right)
\label{sugragaug}
\end{equation}
\begin{equation}
 m_i^{2}\lvert_{anom} = \frac{1}{2}\frac{d \gamma^i}{dt} \left|m_{3/2} - \frac{1}{3}K_l  F^l\right|^2
 \label{sugrascal}
\end{equation}
\begin{equation}
A_{ijk}\lvert_{anom}  = \frac{1}{2}Y^{(0)}_{ijk}\left( \gamma^i + \gamma^j + \gamma^k\right)\left(m_{3/2} - \frac{1}{3}K_l  F^l\right)
\label{sugratril}
\end{equation}
where $K_l = \partial_l K$ and $F^l$are the F-terms. Notice that, as it happens in no-scale models, these contributions to soft terms vanish if  $K_iF^i=3m_{3/2}$.

\subsection{Anomaly mediated soft terms for KKLT and LVS }
It can be seen that the contribution to the scalar masses (\ref{anomscal}) is completely defined in terms of the anomalous dimension. Given that anomalous dimension is coming from the wavefunction renormalisation, the equation (\ref{anomscal}) is telling us that the behaviour of the anomaly mediated contribution to scalars is linked to the behaviour of the renormalisation of the wave-function. This seems to suggest that if one has a K\"ahler potential with no-scale behaviour like
\begin{equation}
 K = -2\log(T+\bar{T} -\phi_0\bar\phi_0)^{3/2}\:,
 \label{noscale}
\end{equation}
the effect of the renormalisation of the wave-functions $\phi_0=\sqrt{Z_0}\phi$ should satisfy the same no-scale behaviour, given that 
\begin{equation}
 K = -2\log(T+\bar{T} -Z_0\phi\bar\phi)^{3/2}\:.
\end{equation}
This would indicate that anomaly mediated contributions to scalar masses follow the same no-scale behaviour as the moduli mediated ones. This no-scale behaviour is produced by the logarithmic structure of the K\"ahler potential. Such a no-scale behaviour is captured by the expression for the scalars in (\ref{sugrascal}).

Concerning the gaugino masses, using (\ref{sugragaug}) one can see that in KKLT
\begin{equation}
M_a^{anom} = -\frac{g_a^2}{16 \pi^2} (3T_G -T_R) \, m_{3/2}\:,
 \label{kkltnoscaleanom1}
\end{equation}
whereas in LVS
\begin{equation}
 M_a^{anom} =  -\frac{g_a^2}{16 \pi^2} (3T_G -T_R) \frac{s^{3/2}\xi}{4\mathcal{V}} \, m_{3/2} \:.
 \end{equation}
By comparing these two expressions we see that in the LVS case there is a  no-scale behaviour  whereas this is not the case in KKLT. That is happening because in LVS, one has
\begin{equation}
  K_{\mathcal{V}} F^{\mathcal{V}} = 3 m_{3/2}
\end{equation}
and this term cancels the $m_{3/2}$ contribution coming from (\ref{anomgaug2}). On the other hand, in KKLT with the nilpotent goldstino, due to the fact that $ K_{X} = 0$ in the vacuum then 
\begin{equation}
  K_{X} F^{X} = 0 
\end{equation}
and there is no such a cancellation. From here we can conclude that anomaly mediation contributions are always subleading in LVS , as
\begin{equation}
 M_a^{anom}\lvert_{LVS} =  -\frac{g_a^2}{16 \pi^2} (3T_G -T_R) \frac{(M_a)_{LVS}}{3}\:.
 \label{lvsnoscaleanom1}
\end{equation}

With respect to the scalar masses, the dominating term for KKLT is
\begin{equation}
m_i^{2}\lvert_{anom}\ = \frac{ \sum_a g_a^4 C_a(i)}{(16 \pi^2)^2} (3T_G -T_R)\  m_{3/2}^2 \:,
 \label{kkltnoscaleanom2}
\end{equation}
whereas for LVS, given the no-scale behaviour at tree level,
\begin{equation}
m_i^{2}\lvert_{anom}\ = \frac{ \sum_a g_a^4 C_a(i)}{(16 \pi^2)^2} (3T_G -T_R)\  \frac{5}{8} \frac{s^{3/2} \xi}{\mathcal{V}}\, m_{3/2}^2 \:.
\end{equation}
We see again here how  the no scale feature of LVS protects it from any contribution coming from anomaly mediation given that 
\begin{equation}
m_i^{2}\lvert_{anom}^{LVS} = \frac{ \sum_a g_a^4 C_a(i)}{(16 \pi^2)^2} (3T_G -T_R)\  (m^2)_{LVS}\:,
 \label{lvsnoscaleanom2}
\end{equation}
whereas for KKLT the anomaly mediation contribution will compete with the one coming from moduli mediation. 

Finally the trilinears satisfy the same pattern as for gauginos and scalars in the LVS case 
\begin{equation}
A_{ijk}\lvert_{anom}^{LVS} =  Y^{(0)}_{ijk}  \sum_{m=i,j,k}\frac{\sum_a g^2_a C_a(m) }{16 \pi^2}\, (A_{ijk})_{LVS}\:,
\end{equation}
whereas in  KKLT there will be a new competition with the moduli mediated term 
\begin{equation}
A_{ijk}\lvert_{anom}^{KKLT} =  Y^{(0)}_{ijk} \sum_{m=i,j,k}\frac{\sum_a g^2_a C_a(m)}{16 \pi^2} \ m_{3/2} \:.
\end{equation}
One can conclude that in LVS anomaly mediation contributions are completely irrelevant but in KKLT they do play a role.

\begin{table}
\begin{center}
{\tabulinesep=1.4mm

   \begin{tabu}{|c|c|c|}
\hline
 & {\bf KKLT} &  {\bf LVS} \\
 \hline\hline
Soft term &  Anomaly mediation &  Anomaly mediation   \\ \hline \hline

$M_{a}$ & $-\left(\frac{g_a^2\, b_a}{16 \pi^2}  \right)\, m_{3/2}$ &$-\left(\frac{g_a^2\, b_a}{16 \pi^2}  \right)\, (M_{1/2})_{\small{LVS}}$   \\ \hline
$m_i^2$ & $\left(\frac{ \sum_a g_a^4\, C_a(i)\,  b_a}{(16 \pi^2)^2}\right) m_{3/2}^2$ &  $\left(\frac{ \sum_a g_a^4\, C_a(i)\,  b_a}{(16 \pi^2)^2}\right)\, (m_0^2)_{\small{LVS}}$  \\ \hline
$A_{ijk}$ &\small{$Y^{(0)}_{ijk} \displaystyle{ \sum_{m=i,j,k}}\sum_a \frac{C_a(m)}{b_a}$} $\, M_{a} $ & \small{$\left(Y^{(0)}_{ijk} \displaystyle{ \sum_{m=i,j,k}} \frac{\sum_a g^2_a C_a(m)  }{16 \pi^2} \right)\,$} $(A_{ijk})_{{LVS} }$\\ \hline
\hline

\end{tabu}}

\small\caption{Summary of different soft terms generated by anomaly mediation  branes for both KKLT and LVS scenarios. The parameter $b_a$ is defined as $b_a = (3T_G -T_R) = (-33/5,\, -1,\, 3)$. Notice that in the scalars and trilinears we are giving just the dominating contribution coming from the anomalous dimension.    }
\label{Tab:anom}
\end{center}
\end{table}
  
  \section{Soft terms on D7-branes}\label{App:D7-brane}
In this section we will analyse the soft-terms in KKLT and LVS for matter fields placed on D7-branes instead of D3-branes.
As we did for the D3-branes, we first analyse the KKLT case and then we study LVS. In both cases, we will add the nilpotent superfield $X$ describing the presence of an anti-D3-brane. Here we do not study in detail the interaction between the anti-D3-brane and the visible sector D7-branes. The presence of the anti-D3-brane could generate a potential for the deformation moduli of the D7-branes, that would move the D7-brane but generically it will not break the gauge group and the structure of the chiral intersections.

\subsection{KKLT with matter fields on D7-branes}

We assume the parametric effective field theory where the K\"ahler potential is 
\begin{equation}
K = -2\log \left(\mathcal{V} \right)  + \tilde{K}_i\ \phi \bar{\phi} + \tilde{Z}_i\ X \bar{X} + \tilde{H}_i\ \phi \bar{\phi} \ X \bar{X} + ...
\end{equation}
where $\mathcal{V}= \tau^{3/2}$. The matter metric is given by
\begin{equation}
 \tilde{K}_i = \alpha \frac{\tau^{1-\lambda}}{\mathcal{V}^{2/3}} \:,
 \end{equation}
 where $\lambda$ is the modular weight, that can take values $\lambda= 0, 1, 1/2$. These values correspond respectively to brane positions, D3-branes (or its dual Wilson line) and D7-branes. We are interested to the last ones. The matter metric for the nilpotent goldstino is the same as in the former sections:   
\begin{equation}
 \tilde{Z}_i = \frac{\beta}{\mathcal{V}^{2/3}} \:.
 \end{equation}
Concerning the quartic interaction, it will be parametrised as 
\begin{equation}
 \tilde{H}_i = \gamma \frac{\tau^{1-\lambda}}{\mathcal{V}^{4/3}} \:.
\end{equation}
The superpotential is again \eqref{Wnil} 
and the scalar potential can then be written as 
\begin{equation}
V = \left(V_{KKLT} + V_{up} \right) + \left( \frac{2}{3}\left( V_{KKLT} + V_{up} \right) + \frac{1}{3}V_{up} \left(1-\frac{3\gamma}{\alpha \beta}\right) + m_\lambda   \right) |\hat\phi |^2\:,
\end{equation}
where $m_\lambda$ is a complicated function of the modular weight $\lambda$ and of the scalar fields. For KKLT, at the minimum it takes the form
\begin{equation}
m_\lambda = \frac{W_0^2}{2s}\frac{1}{a^2 \mathcal{V}^{10/3}}(1-\lambda) \:.
\end{equation}
Notice that unlike the D3-brane case the effective K\"ahler potential cannot be put into the logarithmic form for any values of the parameters $\alpha$, $\beta$ and $\gamma$. The non-zero term
\begin{equation}
\left.\frac{1}{3}V_{up} \left(1-\frac{3\gamma}{\alpha \beta}\right)\right|_{\rm min} = \frac{W_0^2}{2s\mathcal{V}^2} \left(1 - \frac{1}{a^2 \mathcal{V}^{4/3}}   \right) \left(1-\frac{3\gamma}{\alpha \beta}\right) 
\end{equation}
will contribute to the scalar masses. 

Therefore the soft terms for scalar masses can be written in terms of the gravitino mass as
\begin{equation}
m^2 = \left(1-\frac{3\gamma}{\alpha \beta}\right) \, m_{3/2}^2 - \left(\lambda-\frac{3\gamma}{\alpha \beta}\right)\frac{1}{a^2 \mathcal{V}^{4/3}} \, m_{3/2}^2 
\label{ese}
\end{equation}
where the case $\lambda=1/2$ corresponds to D7-branes.\footnote{Notice that for $\lambda=1$  we recover the D3-brane case.}

If we include the $\alpha'$ corrections like in Section \ref{KKLTalphaCorr}, then there is a new term which dominates over the second term in (\ref{ese}) such that 
\begin{equation}
m^2 = \left(1-\frac{3\gamma}{\alpha \beta}\right) \, m_{3/2}^2 + \frac{s^{3/2}\xi}{\mathcal{V}}\frac{3\gamma}{\alpha \beta} (\gamma_1 -\alpha_1 -\beta_1) \, m_{3/2}^2 \:.
\end{equation}
Notice that the prefactor $\left(1-\frac{3\gamma}{\alpha \beta}\right)$ can very easily generate a tachyon. Interestingly if $\gamma=\frac{\alpha \beta}{ 3}$ then the leading contribution to the scalar masses will be given by the $\alpha'$ corrections:
\begin{equation}
m^2 = \frac{s^{3/2}\xi}{\mathcal{V}}(\gamma_1 -\alpha_1 -\beta_1) \, m_{3/2}^2 \:.
\end{equation}

The gaugino masses are dominated by $F^T$, since the gauge kinetic functions is $f=T$. Hence
\begin{equation}
M= \pm \frac{1}{a \mathcal{V}^{2/3}}\, m_{3/2}\:,
\end{equation}
where the relative sign $\pm$ refers to the choice of $W_0\gtrless 0$. Finally the trilinears can be written in terms of the gaugino masses as
\begin{equation}
A_{ijk} = - \frac{3}{2}(2\lambda-1 - s\partial_s \log Y_{ijk}^{(0)}) M \:,
\end{equation}
where in the case of D7-branes one should use $\lambda=1/2$.

\subsubsection*{Cosmological and phenomenological observations}

The discussion for the scalar masses is similar to the one presented in Section~\ref{Sec:KKLTpheno}, for the case when the {\it log hypothesis} is not fulfilled.

The anomaly mediated contributions together with the gaugino masses for D7-branes generate the following competition
\begin{equation}
 M_a^{KKLT} =  \left(\pm \frac{1}{a\mathcal{V}^{2/3}} -\frac{g_a^2}{16 \pi^2} (3T_G -T_R) \right)  \,  m_{3/2}
\end{equation}
and using the same strategy as in \ref{Sec:CosmoPhenoObserv}, one can see that the parameter $\hat{\alpha}$ from \cite{cfno,cfno2,cfno3} is this time $\hat{\alpha}=1$. However, this time the mirage scale is given by
\begin{equation}
M_{mir}=M_{GUT}\, e^{-\frac{100 \pi}{N}} \:.
\label{mirage4}
\end{equation}
Hence, in this case, for $N\sim 11$ one could obtain a TeV mirage scale with a compressed spectrum scenario. For $N<11$ anomaly mediation dominates. The collider phenomenology is similar to the one described in  \ref{Sec:CosmoPhenoObserv} for LVS.  Regarding the KKLT scalar masses, the anomaly mediation terms are suppressed by the loop factor compared to the leading contribution $\left(1-\frac{3\gamma}{\alpha \beta}\right) \, m_{3/2}^2$.

\subsection{LVS with matter fields in D7}

We now study a visible sector realised on D7-branes wrapping a small cycle, i.e. a four-cycle whose volume is (proportional to) $\tau_s$ in the Large Volume Scenario. This can be realised whether the D7-brane cycle is $D_s$ or whether there is a linear relation between the volumes of the two. The first possibility leads to difficulties in allowing an MSSM chiral spectrum on the D7-brane and at the same time having a non-perturbative effect contributing to the superpotential (see \cite{bckmq}). The second situation may be forced by fixing the relation between the two K\"ahler moduli at higher energies (see \cite{Braun:2015pza} for an example).
Here we assume that this is possible.

In this case the K\"ahler potential will be described by
\begin{equation}
K = -2\log \left(\mathcal{V} -\hat\xi \right)  + \tilde{K}_i\ \phi \bar{\phi} + \tilde{Z}_i\ X \bar{X} + \tilde{H}_i\ \phi \bar{\phi} \ X \bar{X} + ...
\end{equation}
where $\mathcal{V}= \tau_b^{3/2}-\tau_s^{3/2}$ and where 
\begin{equation}
 \tilde{K}_i = \alpha \frac{\tau_s^{1-\lambda}}{\mathcal{V}^{2/3}}\,,  \qquad 
 \tilde{Z}_i = \frac{\beta}{\mathcal{V}^{2/3}}\,, \qquad
 \tilde{H}_i = \gamma \frac{\tau_s^{1-\lambda}}{\mathcal{V}^{4/3}} \:.
\end{equation}
 Like in KKLT, the scalar potential can be generically written as
 \begin{equation}
V = \left(V_{LVS} + V_{\alpha'up} \right) + \left( \frac{2}{3}\left( V_{LVS} + V_{\alpha'up} \right) + \frac{1}{3}V_{\alpha'up} \left(1-\frac{3\gamma}{\alpha \beta}\right) + m_\lambda   \right) |\hat\phi |^2 \:,
\end{equation}
where at the LVS minimum $m_\lambda$ takes the form
\begin{equation}
m_\lambda = \frac{9 (1-\lambda)}{(4a_s \tau_s - 1)^2} \frac{W_0^2}{2s\mathcal{V}^2} = \frac{9 (1-\lambda)}{(4a_s \tau_s - 1)^2} \, m_{3/2}^2 
\end{equation}
and where $
\left.\frac{1}{3}V_{\alpha'up} \left(1-\frac{3\gamma}{\alpha \beta}\right) \right|_{\rm min}
$
is subleading, as the Minkowski/dS condition forces
\begin{equation}
V_{\alpha'up} \sim \mathrm{CosmConst_{LVS}} \sim \frac{m_{3/2}^2}{\mathcal{V}}\:,
\end{equation}
where $\mathrm{CosmConst_{LVS}}$ is the absolute value of the LVS AdS cosmological constant when the uplift term is absent (i.e. $\rho=0$).
Hence, the scalar masses at the dS minimum are given by
\begin{equation}
m_0^2 =  \frac{9 (1-\lambda)}{(4a_s \tau_s - 1)^2} \, m_{3/2}^2\:.
\end{equation}
Concerning the gaugino masses, the gauge kinetic function is $f=T_s$ and hence they are dominated by the $F^{T_s}$:
\begin{equation}
M_{1/2} = \pm\frac{3}{4a_s \tau_s - 1}  \, m_{3/2}\:,
\end{equation}
where the relative sign $\pm$ refers to the choice of $W_0\gtrless 0$. Notice that the relation between the scalars and the gauginos is given by
\begin{equation}
m_0^2 = (1-\lambda) M_{1/2}^2\:.
\end{equation}
Finally the trilinears can be written as
\begin{equation}
A_{ijk} = - 3(1-\lambda) M_{1/2}\:.
\end{equation}
For the case of D7-branes, $\lambda=1/2$ and hence
\begin{equation}
m_0^2 = \frac{1}{2} M_{1/2}^2 \qquad\mbox{and}\qquad
A_{ijk} = -\frac{3}{2} M_{1/2}\:.
\end{equation}

\begin{table}
\begin{center}
{\tabulinesep=1.4mm
   \begin{tabu}{|c|c|c|}
\hline
 & {\bf KKLT} &  {\bf LVS} \\
 \hline\hline
Soft term &  D7 &  D7   \\ \hline \hline
$M_{1/2}$ & $\pm\left(\frac{1}{a\vo^{2/3}}  \right) m_{3/2}$ & $\pm\left(\frac{3}{4a\tau_s}\right) m_{3/2}$   \\ \hline
$m_0^2$ & $\left(1-3\omega\right) m_{3/2}^2$ &  $\left(\frac{9(1-\lambda)}{16a^2\tau_s^2}\right)\, m_{3/2}^2$  \\ \hline
$A_{ijk}$ & $\frac{3}{2}(2\lambda-1-s\partial_s\log Y_{ijk})\, M_{1/2} $& $-3(1-\lambda)\, M_{1/2} $\\ \hline
\hline

\end{tabu}}
\small\caption{Summary of different soft terms for the visible sector on  D7-branes for both KKLT and LVS scenarios. Here $\omega=\frac{\gamma_0}{\alpha_0\beta_0}$. Also the modular weight $\lambda$ is kept explicitly with values $\lambda=1/2$ for D7-branes simplifying the expressions. 
}
\label{Tab:D7}
\end{center}
\end{table}

\subsubsection*{Cosmological and phenomenological observations}
The mass of the lightest modulus is
\begin{equation}
 m_{\mathcal{V}}^2 = 5a_s \tau_s \,\frac{s^{3/2}\xi}{ \mathcal{V}}\, m_0^2\:.
\end{equation}
One can see that  in order to avoid the cosmological moduli problem, the bound is
\begin{equation}
 m_0 \gtrsim 10^3\ \mathrm{TeV}\:.
\end{equation}
In this scenario, the gauginos are of the same order as the scalars. Hence all the sparticles are at  $M_{SUSY}\gtrsim 10^3$ TeV. The higgsinos will be of the order $\mu\sim 10$ TeV (if one is able to saturate the last bound) due to the one loop mass contribution induced by the bino and the wino. Therefore, this scenario would need of R-parity violation to avoid dark matter overproduction, and non of the sparticles would be detectable at LHC or at direct or indirect detection experiments.

\appendix

\end{document}